\documentclass[12pt]{iopart}

\usepackage{iopams}
\usepackage{graphicx}

\newcommand{\mean}[1]{\langle{#1}\rangle}
\newcommand{\pro}[2]{\langle{#1}|{#2}\rangle}
\newcommand{\bra}[1]{\langle{#1}|}
\newcommand{\ket}[1]{|{#1}\rangle}

\newcommand{\dgg}{^{\dagger}}

\begin{document}

%%%%%%%%%%%%%%%%%%%%%%%%%%%%%%%%%%%%%%%%%%%%%
%%%%%%%%%%%%%%%%%%%%%%%%%%%%%%%%%%%%%%%%%%%%%
%%%%%%%%%%%%%%%%%%%%%%%%%%%%%%%%%%%%%%%%%%%%%

\title[]
{Zero-dynamics principle for perfect quantum memory in linear networks}

\author{Naoki Yamamoto$\mbox{}^1$ and Matthew R. James$\mbox{}^{2,3}$}

\address{
$\mbox{}^1$ Department of Applied Physics and Physico-Informatics, 
Keio University, Hiyoshi 3-14-1, Kohoku, Yokohama 223-8522, Japan \\
$\mbox{}^2$ Research School of Engineering, 
The Australian National University, Canberra, ACT 0200, Australia \\
$\mbox{}^3$ ARC Centre for Quantum Computation and Communication 
Technology, The Australian National University, ACT 0200, Australia}

\ead{yamamoto@appi.keio.ac.jp, Matthew.James@anu.edu.au}

%%%%%%%%%%%%%%%%%%%%%%%%%%%%%%%%%%%%%%%%%%%%
%%%%%%%%%%%%%%%%%%%%%%%%%%%%%%%%%%%%%%%%%%%%
%%%%%%%%%%%%%%%%%%%%%%%%%%%%%%%%%%%%%%%%%%%%

\begin{abstract}

In this paper, we study a general linear networked system that contains 
a tunable memory subsystem; 
that is, it is decoupled from an optical field for state transportation during 
the storage process, while it couples to the field during the writing or 
reading process. 
The input is given by a single photon state or a coherent state in a pulsed 
light field. 
We then completely and explicitly characterize the condition required on 
the pulse shape achieving the perfect state transfer from the light field 
to the memory subsystem. 
The key idea to obtain this result is the use of zero-dynamics principle, 
which in our case means that, for perfect state transfer, the output field 
during the writing process must be a vacuum. 
A useful interpretation of the result in terms of the transfer function is 
also given. 
Moreover, a four-nodes network composed of atomic ensembles is studied 
as an example, demonstrating how the input field state is transferred to 
the memory subsystem and how the input pulse shape to be engineered 
for perfect memory looks like. 

\end{abstract}

%\pacs{03.67.-a, 03.65.Yz, 02.30.Yy, 42.50.Lc}
%\submitto{\NJP}

%\maketitle

%%%%%%%%%%%%%%%%%%%%%%%%%%%%%%%%%%%%%%%%%%%%
%%%%%%%%%%%%%%%%%%%%%%%%%%%%%%%%%%%%%%%%%%%%
%%%%%%%%%%%%%%%%%%%%%%%%%%%%%%%%%%%%%%%%%%%%

\section{Introduction}

Quantum memory is, in a wide sense, a device that stores or freezes 
a quantum state both spatially and in time. 
A highly successful example is that a light pulse is frozen in a cloud 
of atoms 
\cite{Phillips2001,Hau2001,Polzik 2004,Kuzmich 2005,Sellars 2010}. 
Also quantum memory is of technological importance particularly in 
quantum information science, such as the quantum repeater for 
quantum communication \cite{Repeater 1,Duan 2001,Repeater review}. 
Because of these scientific/technological importance, the field of quantum 
memory has experienced significant progress in both theory and 
experiment. 
We refer to for instance \cite{Lvovsky 2009,Special issue,Focus on} 
for reviewing the current situation of this very active research area.

\begin{figure}
\centering
\includegraphics[scale=0.55]{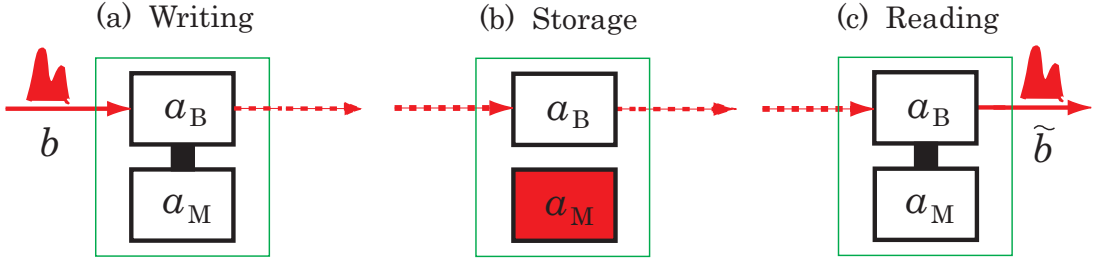}
\caption{
\label{DF memory schematic}
Basic schematic of an ideal quantum memory. 
$a_{\rm M}$ denotes the mode of the memory subsystem used for the 
storage, while $a_{\rm B}$ is the mode of the buffer subsystem, which 
transports the input state to (from) the memory subsystem from (to) 
the optical field with mode $b$ ($\tilde{b}$). 
The system structure can be switched from the stage (a) to (b), or 
from (b) to (c), by tuning some controllable parameters. 
In the stage (b), $a_{\rm M}$ is decoupled from $a_{\rm B}$ and thus 
the optical field, implying that $a_{\rm M}$ is decoherence free. 
In the stages (a) and (c), on the other hand, $a_{\rm M}$ couples to the 
field for state transportation. 
}
\end{figure}

Now let us see a basic and general schematic of an ideal quantum memory 
in an abstract way. 
First of all, we assume that the system contains a subsystem that can 
store any quantum state without loss in a long period of time; 
we call this specific component the {\it memory subsystem}. 
Ideally, the memory subsystem should be completely decoupled from the 
other system components and surrounding environment during it stores 
the state; 
i.e., it is {\it decoherence free (DF)} \cite{KnightNJP2000,Wineland2001,
Lidar2003,BaconPRA2006}. 
Note that, in this storage stage, the memory subsystem is decoupled even 
from the channel used for transferring an input state or retrieving the 
stored state. 
Hence, the second assumption is that, during the writing/reading process, 
the system can be tuned so that the memory subsystem couples to that 
transportation channel. 
That is, the system should be the one that contains a tunable port 
switching the opening/closing of the memory subsystem. 
Indeed this basic schematic is employed in each specific memory device. 
In the case of atoms based on the electromagnetic induced transparency 
(EIT) effect, an isolated memory subsystem is served by a set of 
metastable collective atomic states, and an external control field (with 
Rabi frequency $\omega(t)$) can switch ON/OFF of the coupling between 
the metastable states and the optical field for state transportation 
\cite{Phillips2001,Hau2001,Polzik 2004,Kuzmich 2005,Fleischhauer,
Chen PRL 2013}. 
We also find successful demonstrations in optical cavity or optomechanical 
oscillator arrays 
\cite{LipsonNatPhys2007,Noda 2007,PrebleOptExpress2010,Painter 2011,
Furusawa 2013}, where the switching mechanism is served by adiabatic 
frequency detuning of the memory subsystem. 
Further, a similar switching procedure is employed in the photon echo 
quantum memory \cite{Special issue}. 
Note that, if the system does not contain a switching mechanism, it is 
generally impossible to perfectly transfer an unknown state to a memory 
subsystem (i.e. DF subsystem) \cite{Kielpinski 2013}.

The above-mentioned basic schematic for quantum memory is illustrated 
in Fig.~\ref{DF memory schematic}. 
In the writing stage (a), an input state is sent to the system over an input 
channel with mode $b(t)$. 
Let us here assume that, by devising a ``certain nice procedure", all 
the input state is transferred to the memory subsystem with mode 
$a_{\rm M}$. 
We then close the port, and $a_{\rm M}$ becomes decoherence free. 
In this storage stage (b), ideally, the memory subsystem can store 
the state for a long time. 
Finally, in the reading stage (c), by opening again the port we can retrieve 
the state at any later time, which is sent over the output channel with 
mode $\tilde{b}(t)$.

So what is a  ``certain nice procedure" to achieve the best or hopefully 
perfect state transfer? 
For some typical quantum memories such as EIT and off-resonant 
Raman memories, we can explicitly formulate this problem; 
that is, the question is what is the optimal temporal shape of the 
pulsed light field carrying the input state. 
This optimization problem is very important and actually has been 
deeply investigated in several papers; 
for instance, towards the most efficient atomic memory with EIT, 
in \cite{Gorshkov,Novikova 2007,Novikova 2008,Novikova 2008 b} 
the input wave packet as well as the controllable Rabi frequency 
$\omega(t)$ (see the second paragraph) are carefully designed, 
although the method is a heuristic one based on iterative tuning 
of the parameters. 
It is also notable that the so-called {\it rising exponential} type function 
is found as an effective pulse shape \cite{Muschik2006,He2009,
YiminWang2012}. 
This is a pulse with exponential increase of e.g. the probability density 
of photon counting or the amplitude of a coherent field, which can be 
physically implemented \cite{rising exp experiment 1,rising exp 
experiment 2,rising exp experiment 3}. 
Motivated by these results and further the fact that a large-scale 
quantum memory is required for constructing practical quantum 
communication architecture, the following questions naturally arise. 
Is there a pulse shape for achieving perfect state transfer for general 
and large-scale quantum memory networks? 
Is there a general yet simple guideline for synthesizing such desirable 
pulse shape? 
What is the fundamental origin of the rising exponential function as 
a desirable pulse shape? 
Is the rising exponential pulse optimal in a certain sense? 
Also, is it effective even for a large-scale network? 
Solving these problems should accelerate the progress of the quantum 
memory research in its deeper understanding and practical 
implementation.

In this paper, we consider a general {\it passive linear system}, which 
models a wide variety of systems such as optical systems 
\cite{LipsonNatPhys2007,Noda 2007,PrebleOptExpress2010,
Furusawa 2013,GardinerBook,WisemanBook}, 
nano-mechanical oscillators \cite{Painter 2011,LawPRA1995,Chen 2013}, 
vibration mode of a trapped particle \cite{WinelandRMP2003,Polzik2011}, 
and atomic ensembles \cite{Gorshkov,Novikova 2007,Novikova 2008,
Novikova 2008 b,Muschik2006,He2009,Duan,Kuzmich 2006,Parkins2006,
Parkins2007,Ficek2009,PolzikRMP2010}. 
As mentioned before, the system is assumed to contain a tunable memory 
subsystem; 
that is, by tuning a certain parameter, we can switch opening/closing of 
the memory subsystem, which is DF during the storage period. 
Note that in our case this DF subsystem corresponds to a system having 
the so-called {\it dark mode} \cite{Dong2012,Clerk2012}. 
Another assumption is that an unknown input state to be transferred is 
encoded in a continuos-mode single-photon field or a coherent field. 
With these setups, we give answers to all the questions posed in the above 
paragraph. 
The essential idea is the use of {\it zero-dynamics principle}. 
This concept originates from the classical notion of ``zero" of a {\it transfer 
function}, which is a fundamental tool used in systems and control theory 
\cite{Zhou Doyle book}. 
More precisely, for a general linear classical system 
\[
       \dot{x}=Ax + Bu, ~~
       y=Cx + Du, 
\]
where $u$ is the input, $y$ is the output, and $(A,B,C,D)$ are matrices, 
its input-output relation is simply characterized by a transfer function 
$H[s]=D+C(sI-A)^{-1}B$ as $y[s]=H[s]u[s]$ (see Section~4.3 for detailed 
description); 
then, if the input is given by $u(t)=e^{zt}$, where $z$ is a {\it zero} of the 
transfer function (meaning $H[z]=0$), we have $y(t)=H[z]e^{zt} =0$ under 
some additional conditions. 
That is, the {\it zero-dynamics} is a system whose output is always zero. 
In fact, the concept of zero-dynamics is very important in analysis and 
synthesis for even general nonlinear systems \cite{Isidori,Nijmeijer}.

Actually, the zero-dynamics principle described above can be directly 
applied to general quantum memory problem; 
if the state transfer is perfectly carried out, the input field must be 
completely absorbed in the system, and the output field must not contain 
any small pieces of the input field. 
That is, as a principle, the output should be ``zero" during the writing and 
storage stages. 
Surprisingly, this simple zero-dynamics principle leads us to prove, very 
easily, that a rising exponential function is a unique pulse shape achieving 
the perfect state transfer for general (and thus large-scale) linear passive 
networks. 
Moreover, based on this first main result, we give an explicit, simple, and 
general procedure for designing the wave packet carrying an unknown 
state that is as a result perfectly absorbed in the memory subsystem.

This paper is organized as follows. 
Section~2 provides some preliminaries, describing general passive linear 
systems, optical field states, and linear DF subsystems. 
Also the problem is explicitly formulated in Section~2.4. 
In Section~3, we study a simple single-mode oscillator as a memory 
system, to show the fact that a rising exponential function appears 
as a unique pulse shape achieving the perfect state transfer; 
based on this result, the idea of zero-dynamics principle is discussed. 
Section~4 provides our first main result; 
for general passive linear systems, we derive the rising exponential 
pulse from the zero-dynamics principle. 
Also we give a useful interpretation of this fact in terms of the transfer 
function. 
In Section~5, we present our second main result, showing the concrete 
procedure for perfect writing, storage, and reading; 
this explicitly shows the pulse shape to be synthesized for perfect state 
transfer from the optical field to the memory subsystem. 
Section~6 is devoted to derive the time evolution equations of some 
statistical quantities of the dynamics, which is useful for numerical 
simulation. 
In Section~7, we study a linear network composed of atomic ensembles 
trapped in a cavity, which contains a tunable memory subsystem; 
this example shows how the designed pulse actually looks like and how 
the state transfer from the optical field to the atomic ensembles evolves 
in time. 
Section~8 summarize the paper and discusses some future works. 
In Appendix~A, we briefly examine the case of an {\it active} linear 
memory system. 
Appendix~B provides a case study comparing the zero-dynamics 
principle and the so-called {\it dark state principle}.

{\bf Notation: }
We use the following notations: 
for a matrix $A=(a_{ij})$, the symbols $A^\dagger$, $A^\top$, and 
$A^\sharp$ represent its Hermitian conjugate, transpose, and complex 
conjugation in elements of $A$, respectively; i.e., 
$A^\dagger=(a_{ji}^*)$, $A^\top=(a_{ji})$, and $A^\sharp=(a^*_{ij})$. 
For a matrix of operators we use the same notation, in which case $a_{ij}^*$ 
denotes the adjoint to $a_{ij}$. 
For a time-dependent variable $x(t)$, we denote $\dot{x}(t)=dx(t)/dt$.

%%%%%%%%%%%%%%%%%%%%%%%%%%%%%%%%%%%%%%%%%%%%
%%%%%%%%%%%%%%%%%%%%%%%%%%%%%%%%%%%%%%%%%%%%
%%%%%%%%%%%%%%%%%%%%%%%%%%%%%%%%%%%%%%%%%%%%

\section{Preliminaries and problem formulation}

\subsection{Passive linear systems}

In this paper, we study a general linear open system composed of $n$ 
oscillators with mode $a=[a_1, \ldots, a_n]^\top$ that couples to an 
optical field with continuous mode $b(t)$; 
hence they satisfy $[a_i, a^*_j]=\delta_{ij}$ and $[b(s), b^*(t)]=\delta(s-t)$. 
The system is driven by a quadratic Hamiltonian $H=a^\dagger \Omega a$ 
with $\Omega$ an $n$-dimensional Hermitian matrix. 
The system and the field instantaneously couple through the interaction 
Hamiltonian $H_{\rm int}(t) = i[b^*(t)Ca - a^\dagger C^\dagger b(t)]$ 
with $C$ an $n$-dimensional complex {\it row} vector. 
Then the unitary operator 
\begin{equation}
\label{general unitary}
      U(t_0,t)=\overleftarrow{{\rm exp}}
             \Big[-i\int_{t_0}^t \Big(H + H_{\rm int}(s) \Big)ds \Big]
\end{equation}
($t_0$ is the initial time) produces the Heisenberg equations of 
$a_i(t)=U^*(t_0,t)a_i(t_0)U(t_0,t)$ and 
$\tilde{b}(t)=U^*(t_0,t)b(t)U(t_0,t)$ as follows;
\begin{equation}
\label{general linear}
    \dot{a}(t)=Aa(t) - C\dgg b(t),~~~
    \tilde{b}(t)=Ca(t)+b(t), 
\end{equation}
where $a(t)=[a_1(t), \ldots, a_n(t)]^\top$ and $A=-i\Omega-C\dgg C/2$. 
The second term in $A$ stems from the Ito-correction. 
Note also that, due to the ideal Markov property, the optical field $b(t)$ 
does not have its own dynamical time evolution; 
rather $\tilde{b}(t)$ represents the output of the system.

The above open system with input $b(t)$ and output $\tilde{b}(t)$ is 
called the {\it passive linear system} in the sense that it does not contain 
any active component such as an optical parametric amplifier in optics case. 
As mentioned in Section~1, a passive linear system can model a wide variety 
of systems. 
The main reason why we focus on this class of general systems is that 
it preserves the total energy during the interaction between the system 
and the field. 
This implies that the perfect state transfer is equivalent to the perfect 
energy transfer, which as a result allows us to use the zero-dynamics 
principle to characterize the perfect memory; 
this will be discussed in detail in Section~4. 
Also in Appendix~A, we show a brief case study where the system contains 
an active component. 
For a general treatment of passive linear systems, see \cite{WisemanBook,
GoughPRA2008,Guta Yamamoto}; 
the notation used in this paper follow these references, where in 
general $C$ is an $m\times n$ complex matrix representing $m$ input 
optical fields.

%%%%%%%%%%%%%%%%%%%%%%%%%%%%%%%%%%%%%%%%%%%%

\subsection{Input field states}

In this paper, we consider the case where the input is given by a single 
photon state or a coherent state, which is carried by an optical pulse 
field with continuous-mode $b(t)$. 
They are described as follows.

{\bf Single photon field state: } 
The single photon state in a single mode system is, as is well known, produced 
by acting a creation operator $a^*$ to the ground state $\ket{0}$; i.e. 
\[
      \ket{1}=a^*\ket{0}. 
\]
To describe the continuous-mode single photon field state, we define the 
annihilation and creation process operators 
\begin{equation}
\label{B and Bstar}
       B(\xi) = \int_{-\infty}^\infty \xi^*(t)b(t)dt,~~~
       B^*(\xi) = \int_{-\infty}^\infty \xi(t)b^* (t)dt. 
\end{equation}
$\xi(t)$ is an associated function in ${\mathbb C}$, representing the shape 
of the optical pulse field. 
Also $\xi(t)$ satisfies the normalization condition 
$\int_{-\infty}^\infty |\xi(t)|^2 dt=1$. 
Due to this, $B(\xi)$ and $B^*(\xi)$ satisfy the usual CCR; 
$[B(\xi), B^*(\xi)]=1$. 
The single photon field state is, in a similar way as above, produced by 
acting the creation process operator on the vacuum field $\ket{0}_F$ 
as follows \cite{Zoller 1998,Milburn2008,Munro 2010,Baragiola,Guofeng2013}: 
\begin{equation}
\label{single photon}
   \ket{1_\xi}_F = B^*(\xi)\ket{0}_F 
           = \int_{-\infty}^\infty \xi(t)b^* (t)dt\ket{0}_F.
\end{equation}
Due to the normalization condition of $\xi(t)$, we find that 
$\mbox{}_F\pro{1_\xi}{1_\xi}_F=1$. 
Also note the relation $\mbox{}_F\bra{1_\xi}b^*(t)b(t)\ket{1_\xi}_F
=|\xi(t)|^2$; 
thus, $\xi(t)$ has the meaning of the wave function with $|\xi(t)|^2$ 
the probability of photo detection per unit time. 
Let us now assume that the pulse shape $\xi(t)$ can be expanded as 
\begin{equation}
\label{single photon expansion}
   \xi(t)=\sum_{k=1}^n s_k \gamma_k(t). 
\end{equation}
The coefficient $s_k\in{\mathbb C}$ represents (unknown) classical 
information encoded in the optical field, which satisfies 
$\sum_k|s_k|^2=1$, and $\{\gamma_k(t)\}_{k=1,\ldots,n}$ is a set 
of orthonormal functions satisfying 
\begin{equation}
\label{orthogonal fn}
    \int_{-\infty}^\infty \gamma_j^*(t) \gamma_k(t)dt=\delta_{jk}. 
\end{equation}
Note that $n$ is the number of modes of the system. 
Then the single photon field state \eref{single photon} can be written as 
\begin{equation}
\label{single photon input}
\hspace{-1cm}
    \ket{1_\xi}_F
        =\sum_{k=1}^n s_k \int_{-\infty}^\infty 
               \gamma_k(t) b^*(t)dt \ket{0}_F
        =\sum_{k=1}^n s_k B^*(\gamma_k)\ket{0}_F
        =\sum_{k=1}^n s_k \ket{1_{\gamma_k}}_F. 
\end{equation}
$\ket{1_{\gamma_k}}_F$ is called the single photon code state with 
pulse shape $\gamma_k(t)$ \cite{Milburn2008}; 
from the condition \eref{orthogonal fn}, they are orthonormal, 
i.e. $\mbox{}_F\pro{1_{\gamma_j}}{1_{\gamma_k}}_F=\delta_{jk}$. 
Also the field operator $B(\gamma_k)$ satisfies the CCR 
$[B(\gamma_j), B^*(\gamma_k)]=\delta_{jk}$.

{\bf Coherent field state: } 
Another important state is a coherent state. 
A coherent state in a single mode system is generated by acting a 
displacement operator $e^{\alpha a^* - \alpha^* a}$ on the ground state 
as follows; 
\[
      \ket{\alpha} = e^{\alpha a^* - \alpha^* a} \ket{0}, 
\]
where $\alpha\in{\mathbb C}$ denotes the amplitude of $\ket{\alpha}$. 
Likewise, a coherent field state is defined in terms of the creation and 
annihilation process operators as follows; 
\[
      \ket{f}_F = e^{B^*(f) - B(f)}\ket{0}_F
                      = {\rm exp}\Big[ \int_{-\infty}^\infty 
                               \Big(f(t)b^* (t) - f^*(t)b(t) \Big)dt \Big] 
                                       \ket{0}_F, 
\]
where $f(t)$ is a complex-valued function, representing the amplitude of 
the state; 
that is, this is a coherent pulse field modulated with envelope function $f(t)$. 
Note that $f(t)$ is not necessarily normalized, but its power 
$\int_{-\infty}^\infty |f(t)|^2 dt$ is finite. 
Now we assume that $f(t)$ is given, in terms of the orthonormal functions 
$\{\gamma_k(t)\}_{k=1,\ldots,n}$, by 
\begin{equation}
\label{coherent expansion}
       f(t)=\sum_{k=1}^n \alpha_k \gamma_k(t), 
\end{equation}
where $\alpha_k\in{\mathbb C}$ represents (unknown) classical 
information to be stored. 
The power of $\ket{f}_F$, i.e. the mean photon number in unit time, is then 
given by $\int_{-\infty}^\infty |f(t)|^2 dt=\sum_k|\alpha_k|^2$. 
The coherent field state is as a result described by 
\begin{equation}
\label{coherent input}
\hspace{-6em}
      \ket{f}_F 
           = e^{\sum_k \alpha_k B^*(\gamma_k) - \alpha_k^* B(\gamma_k)}
                                 \ket{0}_F
           = {\rm exp}\Big[ \sum_{k=1}^n \int_{-\infty}^\infty
                               \Big(\alpha_k \gamma_k(t)b^* (t) 
                                              - \alpha_k^* \gamma_k^*(t)b(t) \Big)dt \Big] 
                                       \ket{0}_F.
\end{equation}
Note that this is not a superposition of the coherent field states 
$\ket{\gamma_k}_F$, unlike the single photon field state 
\eref{single photon input}.

%%%%%%%%%%%%%%%%%%%%%%%%%%%%%%%%%%%%%%%%%%%%

\subsection{Decoherence-free subsystem as a memory}

Let us reconsider the linear system \eref{general linear}, which is 
composed of $n$ oscillators. 
Note again that this is an open system with $b(t)$ representing the 
environment field. 
Therefore, during the system works as a memory, ideally some of its 
component, 
the {\it memory subsystem} with mode 
$a_{\rm M}=[a_{m+1}, \ldots, a_n]^\top$, must be decoupled from the 
field $b(t)$; 
this means that the memory subsystem is exactly a decoherence-free 
subsystem \cite{KnightNJP2000,Wineland2001,Lidar2003,BaconPRA2006}. 
But the other component, the {\it buffer subsystem} with mode 
$a_{\rm B}=[a_1 \ldots, a_m]^\top$, still couples to $b(t)$. 
In contrast to $a_{\rm M}$, the state of the buffer subsystem decoheres 
due to the coupling to $b(t)$. 
As a result, in the storage stage, the dynamical equation of the system, 
Eq.~\eref{general linear}, should be of the form 
\begin{equation}
\label{general DF system}
\hspace{-2.2cm}
    \frac{d}{dt}
     \left[ \begin{array}{c}
           a_{\rm B}(t)  \\
           a_{\rm M}(t) \\
          \end{array} \right]
    =\left[ \begin{array}{cc}
           A_{\rm B} & O \\
           O & O \\
          \end{array} \right]
     \left[ \begin{array}{c}
           a_{\rm B}(t)  \\
           a_{\rm M}(t) \\
          \end{array} \right]
    -\left[ \begin{array}{c}
           C_{\rm B}\dgg  \\
           O \\
          \end{array} \right]
          b(t),~~~
     \tilde{b}(t)=C_{\rm B}a_{\rm B}(t)+b(t).
\end{equation}
This equation clearly shows that $a_{\rm M}(t)$ is decoherence free, and 
its state is preserved. 
Note that $a_{\rm M}(t)$ does not appear in the output equation, implying 
that the energy contained in the memory subsystem does not leak out 
into the field.

The general theory of DF subsystems states that, if a DF subsystem exists, 
then the system Hilbert space ${\cal H}$ is decomposed to 
${\cal H}=({\cal H}_1\otimes{\cal H}_2)\oplus{\cal H}_3$, 
where any observable in ${\cal H}_2$ evolves unitarily. 
In our case, the decomposition is of the form 
${\cal H}={\cal H}_{\rm B}\otimes{\cal H}_{\rm M}$; 
thus the system variables are, more precisely, represented by 
$a_{\rm B}\otimes I$ and $I \otimes a_{\rm M}$. 
This special class of continuous-variable DF subsystems appears in 
several situation \cite{Dong2012,Clerk2012,PrauznerJPA2004,Huang2013,
Zambrini2013}. 
Also for a general theory of the linear DF subsystem, including a necessary 
and sufficient condition for a given linear system to have a DF mode, 
see \cite{YamamotoDFS}.

%%%%%%%%%%%%%%%%%%%%%%%%%%%%%%%%%%%%%%%%%%%%

\subsection{Problem description}

Here we describe the problem discussed throughout the paper.

Our system is the general passive linear system \eref{general linear}, and 
it is assumed to be tunable; 
that is, by appropriate tuning of its parameter(s), a part of the system, the 
memory subsystem, couples or decouples to the optical field carrying the 
information. 
Hence the memory subsystem can be switched to be a DF or a non-DF 
subsystem. 
As mentioned in Section~2.3, the memory subsystem stores the state 
during it is in the DF mode, while it should be in the non-DF mode 
when we transfer the input state or retrieve the stored state. 
In particular, we assume that the matrix $A = -i\Omega - C^\dagger C/2$ 
in the writing/reading stages is {\it Hurwitz}, i.e., the real parts of all 
the eigenvalues of $A$ are negative; 
as will be shown later, this condition is necessary for perfect state transfer. 
On the other hand, in the storage stage, the dynamical equation takes the 
form \eref{general DF system}, thus $A$ is not Hurwitz.

The field's initial state is given by $\ket{1_\xi}_F$ in the case of single 
photon input field or $\ket{f}_F$ in the case of coherent input field. 
The system's initial state $\ket{\phi}_S$ is assumed to be separable, 
hence it is given by 
$\ket{\phi}_S=\ket{\phi_1}_{S_1}\otimes \cdots \otimes\ket{\phi_n}_{S_n}$. 
In particular, we will set it to be the ground state 
$\ket{\phi_i}_{S_i}=\ket{0}_{S_i}$ satisfying $a_i\ket{0}_{S_i}=0$.

At time $t_0$, the system and the field start to interact, via the unitary 
operator $U(t_0, t)$ given in Eq.~\eref{general unitary}. 
The composite state at time $t_1$ is then given by 
$\ket{\Psi(t_1)} = U(t_0, t_1)\ket{\phi}_S\ket{1_\xi}_F$ or 
$\ket{\Psi(t_1)} = U(t_0, t_1)\ket{\phi}_S\ket{f}_F$. 
In this writing stage, the memory subsystem is in the non-DF mode, so it 
couples to the field. 
But once the state transfer has been completed, then we switch the system 
parameters so that the memory subsystem becomes decoherence free, 
and its state is preserved during the storage stage. 
Hence, it would be desirable if the state $\ket{\Psi(t_1)}$ is of the 
separable form 
\[
   \ket{\Psi(t_1)}
     =\ket{\phi'(t_1)}_{\rm B}\ket{\phi''(t_1)}_{\rm M}\ket{\psi(t_1)}_F,
\]
and the memory subsystem's state $\ket{\phi''(t_1)}_{\rm M}$ contains 
the full information of the input field state $\ket{1_\xi}_F$ or $\ket{f}_F$. 
Therefore, our goal is to appropriately synthesize the pulse shape $\xi(t)$ 
or $f(t)$, or more precisely their basis functions 
$\{\gamma_k(t)\}_{k=1,\ldots,n}$, 
so that the above desirable transition from the field's initial state to 
$\ket{\phi''(t_1)}_{\rm M}$ occurs.

Lastly we remark on the switching configuration. 
In general, the system matrices $\Omega$ and $C$ (and thus $A$) can 
change in time (i.e. time varying) in order to realize high quality quantum 
memory. 
For instance in \cite{Gorshkov,Novikova 2007,Novikova 2008,Novikova 2008 b}, 
the authors consider the time varying system matrices depending on the 
control field with frequency $\omega(t)$, which is optimized via a heuristic 
method. 
On the other hand, in this paper, we assume that $\Omega$ and $C$ 
are time varying, but they are constant in each stage of the memory procedure; 
that is, they are {\it piecewise constant}. 
In particular, we will take the same system matrices in the writing and 
reading stages.

%%%%%%%%%%%%%%%%%%%%%%%%%%%%%%%%%%%%%%%%%%%%
%%%%%%%%%%%%%%%%%%%%%%%%%%%%%%%%%%%%%%%%%%%%
%%%%%%%%%%%%%%%%%%%%%%%%%%%%%%%%%%%%%%%%%%%%

\section{Perfect state transfer in a single-mode passive linear system}

In this section, we examine a simple case where the memory system is 
given by a single-mode passive linear system. 
As will be shown later, this system does not contain a tunable DF 
component, so it does not work as a perfect storage device. 
Rather the purpose here is that, by focusing only on the writing stage, 
requiring perfect state transfer uniquely determines the pulse shape 
of the input optical field. 
Based on this result, we then derive the explicit form of the output field, 
and show the notion of zero-dynamics in this case. 
Here we study only the single-photon input case, but it is straightforward 
to obtain a similar result in the case of coherent field state.

%%%%%%%%%%%%%%%%%%%%%%%%%%%%%%%%%%%%%%%%%%%%

\subsection{Pulse shaping for perfect state transfer}

Let us consider the following single-mode (i.e. $n=1$) linear system 
interacting with an optical field, which is obtained by setting 
$\Omega=0$ and $C=\sqrt{\kappa}$ in Eq.~\eref{general linear}: 
\begin{equation}
\label{1 dim system}
    \dot{a}(t)=-\frac{\kappa}{2}a(t) -\sqrt{\kappa}b(t),~~~
    \tilde{b}(t)=\sqrt{\kappa}a(t)+b(t), 
\end{equation}
where $\kappa$ is the interaction strength; 
in optics case, this system is typically given by an optical cavity 
with $\kappa$ proportional to the transmissivity of the coupling 
mirror. 
The goal is to send a single photon state over the input pulse field and 
write it perfectly down to the system. 
Note that, however, clearly this system does not contain a tunable DF 
component, hence our interest here is only in the state transfer.

The input state is given by Eq.~\eref{single photon input}, which is now 
essentially $\ket{1_\xi}_F=\ket{1_{\gamma_1}}_F$. 
Thus in this case let us take a superposition of the vacuum and the single 
photon field state 
\[
    \alpha \ket{0}_F + \beta \ket{1_\xi}_F, 
\]
where $\alpha, \beta\in{\mathbb C}$ are the encoded (unknown) classical 
information. 
Recall that the system's initial state is set to the ground state $\ket{0}_S$.

The dynamical equation \eref{1 dim system} has the following 
solution: 
\[
    a(t_1) 
     = e^{-\kappa(t_1-t_0)/2}a(t_0)
          - \sqrt{\kappa}e^{-\kappa t_1/2} 
             \int_{t_0}^{t_1} e^{\kappa s/2} b(s)ds, 
\]
where $t=t_1$ is the time we stop the interaction. 
This can be rewritten as 
\begin{equation}
\label{at solution}
    a^*(t_1)
      = e^{-\kappa(t_1-t_0)/2}a^*(t_0)
          + \sqrt{1-e^{-\kappa(t_1-t_0)}}
             \int_{t_0}^{t_1} \nu(s) b^*(s)ds,
\end{equation}
where 
\[
   \nu(t)=-\sqrt{\frac{\kappa}{e^{\kappa t_1}-e^{\kappa t_0}}}
           e^{\kappa t/2}~~~(t_0\leq t \leq t_1),~~~
   \nu(t)=0~~~(t\leq t_0,~t_1\leq t). 
\]
Note that $\int_{-\infty}^\infty |\nu(t)|^2dt
=\int_{t_0}^{t_1} \nu(t)^2dt=1$. 
Equation~\eref{at solution} can be further represented as 
\begin{equation}
\label{mode transfer}
\hspace{-3.9em}
   U^*(t_0, t_1) a^*(t_0) U(t_0, t_1)
    = e^{-\kappa(t_1-t_0)/2}a^*(t_0)\otimes I_F
          + \sqrt{1-e^{-\kappa(t_1-t_0)}} I_S \otimes B^*(\nu).
\end{equation}
$B^*(\nu)$ is the field creation operator with pulse shape $\nu(t)$, which is 
defined in Eq.~\eref{B and Bstar}, and $U(t_0, t_1)$ is the unitary operator 
given in Eq.~\eref{general unitary}. 
From the above equation we find that, in the limit 
$t_0\rightarrow -\infty$, the field creation operator $B^*(\nu)$ 
is completely mapped to the system creation operator $a^*(t_1)$. 
This means that the perfect state transfer from the optical pulse field to 
the system mode can be carried out as shown below. 
In the Schr\"{o}dinger picture, the whole state at time $t=t_1$ is given by 
\begin{eqnarray}
& & \hspace*{-4.5em}
    \ket{\Psi(t_1)}
      = U(t_0, t_1)
         \ket{0}_S (\alpha \ket{0}_F + \beta \ket{1_\xi}_F)
      = U(t_0, t_1) 
          \Big[\alpha I_S\otimes I_F + \beta I_S\otimes B^*(\xi)\Big]
            \ket{0}_S\ket{0}_F
\nonumber \\ & & \hspace*{-1.5em}
      = U(t_0, t_1) 
         \Big[ \alpha I_S\otimes I_F + \beta I_S\otimes B^*(\xi)\Big]
           U^*(t_0, t_1) U(t_0, t_1) \ket{0}_S\ket{0}_F
\nonumber \\ & & \hspace*{-1.5em}
      = \Big[ \alpha I_S\otimes I_F 
           + \beta U(t_0, t_1)(I_S\otimes B^*(\xi))U^*(t_0, t_1) \Big]
              \ket{0}_S\ket{0}_F, 
\nonumber
\end{eqnarray}
where in the last equality $U(t_0, t_1) \ket{0}_S\ket{0}_F=
\ket{0}_S\ket{0}_F$ is used. 
Let us now set the input pulse shape to be $\xi(t)=\nu(t)$. 
Then, from Eq.~\eref{mode transfer}, we have 
\begin{eqnarray}
& & \hspace*{-4.5em}
    \ket{\Psi(t_1)}
     = \Big[ \alpha I_S\otimes I_F 
        + \frac{\beta}{\sqrt{1-e^{-\kappa(t_1-t_0)}}} a^*(t_0)\otimes I_F
\nonumber \\ & & \hspace*{2em}
    \mbox{}
     - \frac{\beta e^{-\kappa(t_1-t_0)/2}}{\sqrt{1-e^{-\kappa(t_1-t_0)}}}
          U(t_0, t_1) a^*(t_0) U^*(t_0, t_1) \Big] \ket{0}_S\ket{0}_F
\nonumber \\ & & \hspace*{-1.6em}
    = \Big[ \alpha \ket{0}_S 
          + \frac{\beta}{\sqrt{1-e^{-\kappa(t_1-t_0)}}} \ket{1}_S
             \Big] \otimes \ket{0}_F
\nonumber \\ & & \hspace*{2em}
    \mbox{}
     - \frac{\beta e^{-\kappa(t_1-t_0)/2}}{\sqrt{1-e^{-\kappa(t_1-t_0)}}}
          U(t_0, t_1) a^*(t_0) U^*(t_0, t_1) \ket{0}_S\ket{0}_F. 
\nonumber
\end{eqnarray}
Therefore, in the limit $t_0\rightarrow -\infty$, we have 
\[
    \ket{\Psi(t_1)}
     = \Big[ \alpha \ket{0}_S + \beta \ket{1}_S \Big] \otimes \ket{0}_F, 
\]
which means that the input field state is completely transferred to the 
system state. 
In particular, in the case $t_0\rightarrow -\infty$ and $t_1=0$, 
the input pulse shape is given by 
\begin{equation}
\label{RE example}
   \xi(t)=-\sqrt{\kappa}e^{\kappa t/2}~~~(t \leq 0),~~~
   \xi(t)=0~~~(0<t). 
\end{equation}
This is called the {\it rising exponential pulse} 
\cite{Gorshkov,Muschik2006,He2009,YiminWang2012,
rising exp experiment 1,rising exp experiment 2,rising exp experiment 3}. 
It is clear from the above discussion that the rising exponential is the 
unique pulse shape for perfect state transfer from the single photon 
field to the system.

%%%%%%%%%%%%%%%%%%%%%%%%%%%%%%%%%%%%%%%%%

\subsection{Explicit form of the output field}

Let us further study the final state 
$\ket{\Psi(t_1)}
=U(-\infty, t_1)\ket{0}_S(\alpha\ket{0}_F+\beta\ket{1_\xi}_F)$, where 
the input pulse shape is now set to 
\begin{equation}
\label{input pulse}
   \xi(t)=-\sqrt{\gamma}e^{\gamma t/2}~~~(t \leq 0),~~~
   \xi(t)=0~~~(0<t). 
\end{equation}
It is possible to obtain the explicit solution:
\[
    \ket{\Psi(t_1)}=\ket{0}_S\ket{\psi^{(1)}(t_1)}_F
                   + \ket{1}_S\ket{\psi^{(0)}(t_1)}_F, 
\]
where 
\[
\hspace*{-5.5em}
    \ket{\psi^{(1)}(t_1)}_F
     = \alpha \ket{0}_F + \beta\ket{1_\xi}_F 
          - \beta \int_{-\infty}^{t_1} \xi'(s)b^*(s)ds \ket{0}_F,~~~
    \ket{\psi^{(0)}(t_1)}_F
     = -\frac{\beta}{\sqrt{\kappa}}\xi'(t_1)\ket{0}_F,
\]
with 
\[
   \xi'(t)=\frac{-2\kappa\sqrt{\gamma}}{\kappa+\gamma} e^{\gamma t/2}~~~
    (t\leq 0),~~~~
   \xi'(t)=\frac{-2\kappa\sqrt{\gamma}}{\kappa+\gamma} e^{-\kappa t/2}~~~
    (0 < t).
\]

First, at $t_1=0$, we have 
\[
   \ket{\Psi(0)}
    = \ket{0}_S\otimes \Big[
        \alpha\ket{0}_F 
          + \beta \frac{\kappa-\gamma}{\kappa+\gamma} \ket{1_\xi}_F \Big]
     + \frac{2\kappa\beta}{\kappa+\gamma}\sqrt{\frac{\gamma}{\kappa}}
          \ket{1}_S\ket{0}_F, 
\]
which becomes $\ket{\Psi(0)}
= (\alpha \ket{0}_S + \beta \ket{1}_S) \otimes \ket{0}_F$ only when 
$\kappa=\gamma$. 
That is, the frequency bandwidth of the input pulse shape has to 
be exactly equal to that of the memory system to attain the perfect 
state transfer. 
This is a form of the so-called {\it impedance matching} for efficient 
energy transfer; 
in Section 7, we will discuss the matching condition in a more practical 
setup where the system is composed of a cavity and atomic ensembles.

Next, in the limit $t_1\rightarrow\infty$, the whole state again becomes 
separable: 
\[
    \ket{\Psi(\infty)}
      = \ket{0}_S\otimes 
          (\alpha \ket{0}_F + \beta \ket{1_{\tilde{\xi}}}_F), 
\]
where 
\[
   \tilde{\xi}(t)
      =\frac{\kappa-\gamma}{\kappa+\gamma} 
           \sqrt{\gamma}e^{\gamma t/2}~~~
    (t \leq 0),~~~~
   \tilde{\xi}(t)
      =\frac{2\kappa}{\kappa+\gamma} \sqrt{\gamma}e^{-\kappa t/2}~~~
    (0 < t).
\]
This $\tilde{\xi}(t)$ represents the pulse shape of the output optical 
field over the whole period. 
Hence, if $\kappa=\gamma$, the output field is vacuum in the writing 
stage $t\leq 0$; 
this means that the single photon input field is completely absorbed into 
the system, and the output field does not contain any pieces of the input 
state. 
In the optics case where the system is given by a cavity, a physical meaning 
of this fact is that it happens destructive interference between the light 
field reflected at the coupling mirror and the transmissive light field 
leaking from the cavity; 
as a result, the output field of the cavity is always in vacuum, i.e. ``zero", 
while the system's state still dynamically changes in time. 
In general, the dynamics of a system whose output is always zero is called 
the zero dynamics \cite{Zhou Doyle book,Isidori,Nijmeijer}. 
Hence in this case the cavity dynamics during the writing process is 
exactly a zero dynamics.

%%%%%%%%%%%%%%%%%%%%%%%%%%%%%%%%%%%%%%%%%%%%
%%%%%%%%%%%%%%%%%%%%%%%%%%%%%%%%%%%%%%%%%%%%
%%%%%%%%%%%%%%%%%%%%%%%%%%%%%%%%%%%%%%%%%%%%

\section{Zero-dynamics principle for perfect state transfer}

In this section, based on the so-called energy-balanced identity, we show 
the notion of zero-dynamics principle as a guideline for perfect state 
transfer in general passive linear systems. 
Then, we prove that the zero-dynamics principle readily leads to 
the rising exponential function as a unique pulse shape. 
Further, a useful view of the zero-dynamics principle in terms of 
the transfer function is provided.

Note that the zero-dynamics principle is essentially equivalent to the 
so-called {\it dark state principle}; 
this idea was first employed in \cite{Cirac 1997} for the application to 
a lossless node-to-node state transfer in a cavity QED network, and later 
several applications have been developed, e.g. lossless gate operation 
\cite{Beige 2000}. 
Appendix~B provides a detailed case study comparing the zero-dynamics 
principle and the dark state principle and then discusses their difference.

%%%%%%%%%%%%%%%%%%%%%%%%%%%%%%%%%%%%%%%%%%%%

\subsection{Input-output relation of the pulse shape}

First we remark that the pulse shape of the single-photon input field, 
$\xi(t)$, and that of the output field, say $\tilde{\xi}(t)$, can be 
connected through a dynamical equation having the same form as 
Eq.~\eref{general linear}. 
Actually, by multiplying $\ket{\phi}_S\ket{1_\xi}_F$ by 
Eq.~\eref{general linear} from the right hand side and using the 
relation $b(t)\ket{1_\xi}_F=\xi(t)\ket{0}_F$, 
we find that the mean photon number of the output field is given by 
\[
      \tilde{n}(t) = \bra{\phi,1_\xi}\tilde{b}^*(t)\tilde{b}(t)\ket{\phi,1_\xi}
        = \Big| \xi(t) - Ce^{At}\int_{-\infty}^t e^{-As}C^\dagger \xi(s)ds  \Big|^2. 
\]
By definition this should be written as $\tilde{n}(t)=|\tilde{\xi}(t)|^2$, 
hence $\xi(t)$ and $\tilde{\xi}(t)$ are related through the following 
dynamics: 
\begin{equation}
\label{dynamics of xi}
            \dot{\eta}(t)=A\eta(t) - C\dgg \xi(t),~~~
            \tilde{\xi}(t)=C\eta(t)+\xi(t), 
\end{equation}
where $\eta(t)$ is a $n$-dimensional c-number vector. 
Note that $\eta(t)$ does not have a particular physical meaning, unlike 
the vector $m(t)$ appearing just below.

The same classical dynamical equation holds for the case of coherent 
input field; 
noting that $\bra{\phi,f}b(t)\ket{\phi,f}=f(t)$, we readily see that 
the vector of mean amplitudes, 
$m(t)=[\mean{a_1(t)}, \ldots, \mean{a_n(t)}]^\top$ 
with $\mean{a_i(t)}=\bra{\phi,f}a_i(t)\ket{\phi,f}$, follows 
\begin{equation}
\label{dynamics of m}
            \dot{m}(t)=Am(t) - C\dgg f(t),~~~
            \tilde{f}(t)=Cm(t) + f(t), 
\end{equation}
where $\tilde{f}(t)$ is the amplitude of the output field $\tilde{b}(t)$. 
This equation has the same form as Eq.~\eref{dynamics of xi}, hence 
in what follows we use Eq.~\eref{dynamics of xi} when discussing the 
input-output relation of the corresponding wave packets.

%%%%%%%%%%%%%%%%%%%%%%%%%%%%%%%%%%%%%%%%%%%%

\subsection{The zero-dynamics principle and rising exponential pulse}

To pose the zero-dynamics principle, it is important to first remember that, 
for the general passive linear system \eref{general linear}, the following 
{\it energy balance identity} \cite{Hush} holds: 
\begin{equation}
\label{balanced equality}
       \int_{t_0}^t \tilde{b}^*(s)\tilde{b}(s)ds  +  a^\dagger(t) a(t) 
         = \int_{t_0}^t b^*(s)b(s)ds  +  a^\dagger(t_0) a(t_0). 
\end{equation}
This indicates that the total energy contained in the system and the field 
is preserved for all time. 
Indeed, from Eq.~\eref{balanced equality} we immediately have 
\[
        \int_{t_0}^t |\tilde{\xi}(s)|^2ds  +  \mean{a^\dagger(t) a(t)} 
         = \int_{t_0}^t |\xi(s)|^2 ds  +  \mean{a^\dagger(t_0) a(t_0)}, 
\]
where the mean is taken for the state $\ket{\phi,1_\xi}$. 
Now we assume $\mean{a^\dagger(t_0) a(t_0)}=0$. 
Then, for the energy of the input pulse field to be completely transferred to 
the system, we need $\tilde{\xi}(t)=0$ for $\forall t \in[t_0, t_1]$ with 
$t_1$ the stopping time of the writing process. 
This is a rigorous description, in the case of passive linear systems, of 
the zero-dynamics principle; 
that is, for the general quantum memory problem with a passive system, 
the output field must be vacuum (i.e. ``zero") for perfect state transfer. 
Surprisingly, this simple condition uniquely determines the form of the 
input pulse shape $\xi(t)$ as shown below.

First, from the requirement $\tilde{\xi}(t)=C\eta(t)+\xi(t)=0$, we have 
$C^\dagger C\eta(t)+C^\dagger\xi(t)=0$, which further leads to 
\begin{equation}
\label{zero dynamics}
    \dot{\eta}(t)=(A+C^\dagger C)\eta(t)
                 =\Big( -i\Omega +\frac{1}{2}C^\dagger C\Big)\eta(t)
                 =-A^\dagger \eta(t). 
\end{equation}
This has the solution $\eta(t)=e^{-A^\dagger (t-t_1)}\eta_1$, with $\eta_1$ 
a fixed vector. 
Thus, again from the condition $C\eta(t)+\xi(t)=0$, we have 
\[
    \xi(t) = -C\eta(t)
           = -\eta(t)^\top C^\top
           = -\eta_1^\top e^{-A^\sharp (t-t_1)} C^\top. 
\]
Note that the input is sent during the writing stage $t\leq t_1$, and $\xi(t)=0$ 
in the storage and reading stages in $t_1\leq t$. 
Taking this into account, we end up with the expression 
\begin{equation}
\label{RE in section 4}
    \xi(t) = -\eta_1^\top e^{-A^\sharp (t-t_1)} C^\top \Theta(t_1-t), 
\end{equation}
where $\Theta(t)$ is the Heaviside step function taking $1$ for 
$t\geq 0$ and $0$ for $t<0$. 
Since $A$ is Hurwitz, as assumed in Section~2.4, the real parts of all the 
eigenvalues of $-A^\sharp$ are strictly positive. 
Hence Eq.~\eref{RE in section 4} is a generalization of the rising 
exponential function. 
In fact, this immediately recovers the result \eref{RE example} in the 
example, where $A=-\kappa/2$, $C=\sqrt{\kappa}$, and particularly 
$t_1=0$. 
Lastly we remark that the zero dynamics is given by 
Eq.~\eref{zero dynamics}, which is defined up to time $t_1$.

%%%%%%%%%%%%%%%%%%%%%%%%%%%%%%%%%%%%%%%%%%

\subsection{Transfer function approach}

Let us define the (two sided) Laplace transform of a signal $x(t)$ by 
\[
      x[s]=\int_{-\infty}^\infty x(t) e^{-st} dt, ~~~s\in{\mathbb C}. 
\]
Note that, when $s=i\omega~(\omega\in{\mathbb R})$, this represents 
the Fourier transformation. 
Then the Laplace transformation of Eq.~\eref{dynamics of xi} gives 
\begin{equation}
\label{general transfer fn}
    \tilde{\xi}[s] = G[s]\xi[s],~~~
    G[s] = 1-C(sI-A)^{-1}C^\dagger. 
\end{equation}
The {\it transfer function} $G[s]$ characterizes the input-output relation 
of the linear system \eref{dynamics of xi} in the Laplace domain. 
As explained in Section~1, the zero-dynamics principle originates from the 
classical notion of ``zero" of a transfer function \cite{Zhou Doyle book}, and 
we can now explicitly describe this fact.

First, to see the idea let us return to the example studied in Section~3. 
The input pulse shape is given by Eq.~\eref{RE example}, whose 
Laplace transformation is $\xi[s]=\sqrt{\kappa}/(s-\kappa/2)$. 
Also in this case the transfer function is given by 
\begin{equation}
\label{transfer fn example}
      G[s] = 1 - \frac{\kappa}{s+\kappa/2}
              = \frac{s-\kappa/2}{s+\kappa/2}. 
\end{equation}
Hence, the output is computed as 
\[
    \tilde{\xi}[s] 
    = \frac{s-\kappa/2}{s+\kappa/2}\cdot
        \frac{\sqrt{\kappa}}{s-\kappa/2}
    = \frac{\sqrt{\kappa}}{s+\kappa/2}, 
\]
and its inverse Laplace transform then yields 
\[
   \tilde{\xi}(t)=0~~~
    (t \leq 0),~~~~
   \tilde{\xi}(t)
      = \sqrt{\kappa}e^{-\kappa t/2}~~~
    (0 < t).
\]
Thus, we again see that the input field is completely absorbed in the system 
during $t\leq t_1=0$; 
i.e. the perfect state transfer has been carried out. 
The most notable point is clearly that the {\it zero} of $G[s]$ is erased 
(in general, if for a transfer function $H[z]$ there exists a $z$ satisfying 
$H[z]=0$, then $z$ is called a zero).

Now we can generalize the above fact; 
for simplicity, we set $t_1=0$. 
The transfer function of the general passive linear system is given by 
Eq.~\eref{general transfer fn}. 
Also the Laplace transformation of the rising exponential function 
\eref{RE in section 4} is given by 
\[
       \xi[s] = \int_{-\infty}^0 -\eta_1^\top e^{-A^\sharp t} C^\top e^{-st} dt  
                 = C(sI+A^\dagger)^{-1}\eta_1. 
\]
Therefore, the output is computed as 
\begin{eqnarray}
& & \hspace*{-4em}
     \tilde{\xi}[s] = G[s]\xi[s] 
          = \Big[ 1-C(sI-A)^{-1}C^\dagger \Big]C(sI+A^\dagger)^{-1}\eta_1
\nonumber \\ & & \hspace*{-2.5em}
          = C(sI+A^\dagger)^{-1}\eta_1
           - C(sI-A)^{-1} \Big[ (sI-A) - (sI+A^\dagger) \Big](sI+A^\dagger)^{-1}\eta_1
\nonumber \\ & & \hspace*{-2.5em}
          = C(sI-A)^{-1}\eta_1. 
\nonumber
\end{eqnarray}
Since $A$ is Hurwitz, the output $\tilde{\xi}[s]$ does not contain a zero, 
implying $\tilde{\xi}(t)=0,~\forall t\leq 0$.

In general, if a transfer function contains a (transmission) zero, then there 
alway exists an input signal such that the corresponding output takes zero 
\cite{Zhou Doyle book}. 
Therefore, we have the following interpretation of the zero-dynamics 
principle for quantum memory in terms of transfer function; 
in general, a linear memory system needs to have a zero for perfect state 
transfer, and the input state is sent over an optical field whose pulse shape 
is characterized by that zero. 
This view would be useful particularly in the case of multi input channels.

%%%%%%%%%%%%%%%%%%%%%%%%%%%%%%%%%%%%%%%%%%%%
%%%%%%%%%%%%%%%%%%%%%%%%%%%%%%%%%%%%%%%%%%%%
%%%%%%%%%%%%%%%%%%%%%%%%%%%%%%%%%%%%%%%%%%%%

\section{Perfect memory procedure in passive linear system}

In this section, we provide a detailed procedure to achieve the perfect 
memory, which is composed of the following three stages; 
the perfect state transfer from the input field to the memory subsystem 
(writing), decoherence-free preservation of the transferred state (storage), 
and the appropriate retrieving of the system state into the output field 
(reading). 
The setup was described in Section~2.4; 
note again that the system matrices $\Omega$ and $C$ (thus $A$) change 
depending on the memory stage, but they are piecewise constant. 
Also, as motivated by the result obtained in Section~3, we will take 
$t_0\rightarrow -\infty$, while keeping general $t_1$.

One of the main questions is as follows; 
although we have derived the rising exponential function \eref{RE in section 4} 
from the zero-dynamics principle, it still contains some parameters that should 
be chosen appropriately; 
more precisely, it is given by 
$\xi(t) = -\eta_1^\top e^{-A^\sharp (t-t_1)} C^\top \Theta(t_1-t)$, 
and we need to determine $\eta_1$ so that the input field state is 
completely transferred to the memory subsystem. 
In this section, we will see that this synthesis problem is clearly solved.

%%%%%%%%%%%%%%%%%%%%%%%%%%%%%%%%%%%%%%%%%%%%

\subsection{The writing stage}

The solution of the general linear equation \eref{general linear} is 
explicitly given by 
\[
    a(t) 
     = e^{A (t-t_0)}a(t_0)
          - e^{At} \int_{t_0}^t e^{-A s} C\dgg b(s)ds. 
\]
Since $A$ is Hurwitz, we can take the limit $t_0\rightarrow -\infty$, 
which yields 
\[
    a^\sharp (t_1)
     = - e^{A^\sharp t_1} \int_{-\infty}^{t_1} 
          e^{-A^\sharp s} C^\top b^*(s)ds, 
\]
where again $t_1$ is the stopping time of the writing process. 
Let us now define the following vector of rising exponential functions: 
\begin{equation}
\label{optimal pulse shape vector}
    \nu(t) = - e^{-A^\sharp (t-t_1)} C^\top \Theta(t_1-t). 
\end{equation}
Then the above solution of $a^\sharp(t_1)$ can be expressed as 
\[
    a^\sharp(t_1) 
     = \int_{-\infty}^\infty \nu(t) b^*(t)dt
     = [I_S\otimes B^*(\nu_1), \ldots, I_S\otimes B^*(\nu_n)]^\top. 
\]
This is a vector of creation operators $a^*_k(t_1)$, implying 
that it has to satisfy the canonical commutation relation 
$aa\dgg - (a^\sharp a^\top)^\top=I$; 
actually, we have 
\begin{eqnarray}
& & \hspace*{-4em}
    aa\dgg - (a^\sharp a^\top)^\top
     = \int_{-\infty}^{\infty}\int_{-\infty}^{\infty}
        \nu(s)^\sharp [b(s), b^*(\tau)] \nu(\tau)^\top ds d\tau
     = \int_{-\infty}^{\infty}
        \nu(s)^\sharp \nu(s)^\top ds 
\nonumber \\ & & \hspace*{-3em}
     = e^{A t_1} \Big[
         \int_{-\infty}^{t_1} e^{-As}C\dgg C e^{-A\dgg s}ds 
           \Big]e^{A\dgg {t_1}}
     = e^{A t_1} \Big[
         \int_{-\infty}^{t_1} \frac{d}{ds}\Big( e^{-As}e^{-A\dgg s}\Big)ds 
           \Big]e^{A\dgg {t_1}}
     = I. 
\nonumber
\end{eqnarray}
This relation shows that $\nu_i(s)$ are orthonormal; 
$\int_{-\infty}^\infty \nu_i^*(s)\nu_j(s) ds=\delta_{ij}$.

{\bf Case I: Single photon state.} 
We consider the case where the input is the single photon field state 
\eref{single photon input}. 
Also the system is assumed to be in the separable ground state at the initial 
time $t_0$.  
Then, through the interaction (see Fig.~2~(a)) the whole state changes to
\begin{eqnarray}
& & \hspace*{-3em}
    \ket{\Psi(t_1)}
     = U(-\infty, t_1)\ket{0,\ldots,0}_S \ket{1_\xi}_F
     = U(-\infty, t_1)\ket{0,\ldots,0}_S 
                     \otimes \sum_k s_k B^*(\gamma_k)\ket{0}_F
\nonumber \\ & & \hspace*{-0.05em}
     = U(-\infty, t_1)\Big[ \sum_k s_k I_S\otimes B^*(\gamma_k) \Big]
          \ket{0,\ldots,0}_S\ket{0}_F
\nonumber \\ & & \hspace*{-0.05em}
     = \sum_k s_k U(-\infty, t_1)\big[ I_S\otimes B^*(\gamma_k) \big]
         U^*(-\infty, t_1)
           \ket{0,\ldots,0}_S\ket{0}_F,
\nonumber
\end{eqnarray}
where $U(t_0, t_1)$ denotes the unitary time evolution \eref{general unitary}. 
We here set the basis functions $\gamma_k(t)$ to the rising exponential 
functions $\nu_k(t)$ given in Eq.~\eref{optimal pulse shape vector}, meaning 
that the input pulse shape \eref{single photon expansion} is chosen as 
\begin{equation}
\label{xi in section 5}
      \xi(t)=\sum_k s_k \nu_k(t). 
\end{equation}
Then, noting that 
$a_k^*(t_1)=U^*(-\infty, t_1)a_k^*(-\infty)U(-\infty, t_1)=I_S\otimes B^*(\nu_k)$, 
we obtain
\begin{equation}
\label{transfer process}
    \ket{\Psi(t_1)}
     = \sum_k s_k a_k^*(-\infty) \ket{0,\ldots,0}_S\ket{0}_F
     = \Big[ \sum_k s_k \ket{1^{(k)}}_S \Big] \otimes \ket{0}_F, 
\end{equation}
where $\ket{1^{(k)}}_S=\ket{0,\ldots,1,\ldots,0}_S$ with $1$ 
appearing only in the $k$th entry. 
Therefore, through the interaction, at time $t=t_1$ the system 
completely acquires the input code state with coefficient $\{s_k\}$; 
the resulting system state is highly entangled among the nodes 
(see Fig.~2~(b)). 
The optimal input pulse shape is given by the rising exponential 
function of the form \eref{RE in section 4}, as expected in Section~4.2. 
But the point here is that we now know that the parameter vector 
$\eta_1$ in Eq.~\eref{RE in section 4} exactly corresponds to the 
superposition coefficients $\{s_k\}$. 
Together with the structure of the memory subsystem, this fact 
tells us how we should design the input pulse shape $\xi(t)$; 
this will be more precisely discussed in the next subsection.

{\bf Case II: Coherent state.} 
Next, let us consider the case where the input is the coherent field state 
\eref{coherent input}. 
Again the system is in the ground state at $t_0\rightarrow -\infty$. 
Then, through the interaction the whole state becomes 
\begin{eqnarray}
& & \hspace*{0em}
    \ket{\Psi(t_1)}
     = U(-\infty, t_1)\ket{0,\ldots,0}_S 
                     \otimes 
                e^{\sum_k \alpha_k B^*(\gamma_k) - \alpha_k^* B(\gamma_k)}\ket{0}_F\nonumber \\ & & \hspace*{3em}
     = U(-\infty, t_1)
                e^{\sum_k \alpha_k B^*(\gamma_k) - \alpha_k^* B(\gamma_k)}
                      U^*(-\infty, t_1)
                             \ket{0,\ldots,0}_S\ket{0}_F. 
\nonumber
\end{eqnarray}
Therefore, by setting the basis functions $\gamma_k(t)$ to the rising 
exponential \eref{optimal pulse shape vector} i.e. 
\begin{equation}
\label{f in section 5}
      f(t) = \sum_k \alpha_k \nu_k(t) = \alpha^\top \nu(t)
            = - \alpha^\top e^{-A^\sharp (t-t_1)} C^\top \Theta(t_1-t), 
\end{equation}
and again noting that 
$a_k^*(t_1)=U^*(-\infty, t_1)a_k^*(-\infty)U(-\infty, t_1)
=I_S\otimes B^*(\nu_k)$, we obtain
\begin{equation}
\label{transfer process}
    \ket{\Psi(t_1)}
     = e^{\sum_k \alpha_k a_k^*(-\infty) - \alpha_k^* a_k(-\infty)}
                             \ket{0,\ldots,0}_S\ket{0}_F
     = \ket{\alpha_1,\ldots,\alpha_n}_S\ket{0}_F. 
\end{equation}
That is, the system state is changed to the product of coherent states 
$\ket{\alpha_k}$. 
Hence, similar to the single photon case, the perfect state transfer is 
possible by sending the information $\{\alpha_k\}$ over the rising 
exponential pulse field.

%%%%%%%%%%%%%%%%%%%%%%%%%%%%%%%%%%%%%%%%%%

\subsection{The storage stage}

As mentioned in Sections~1 and 2.4, the key architecture of an ideal 
memory device is that the system contains the tunable memory subsystem 
that can be switched to DF mode in the storage stage or non-DF mode 
in the other two stages; 
now we are in the storage stage.
Especially to describe the idea explicitly, let us consider the case $n=5$ 
only in this subsection and assume that, after the writing process has 
been completed at time $t=t_1$, the system can be immediately switched 
so that its dynamical equation is of the following form:
\[
\hspace{-1.5cm}
    \frac{d}{dt}
     \left[ \begin{array}{c}
           a_{\rm B}  \\
           a_{\rm M} \\
          \end{array} \right]
    =\left[ \begin{array}{cc}
           A_{\rm B} & O \\
           O & O  \\
          \end{array} \right]
     \left[ \begin{array}{c}
           a_{\rm B}  \\
           a_{\rm M} \\
          \end{array} \right]
    -\left[ \begin{array}{c}
           C_{\rm B}\dgg  \\
           O \\
          \end{array} \right]
          b(t),~~~
     \tilde{b}(t)=C_{\rm B}a_{\rm B}(t)+b(t),
\]
where $a_{\rm B}=[a_1, a_2]^\top$ is the buffer mode and 
$a_{\rm M}=[a_3, a_4, a_5]^\top$ is the memory mode. 
Clearly, $a_{\rm M}$ constitutes a DF subsystem. 
Hence, in the single photon input case, the whole state 
$\sum_{k=1}^5 s_k \ket{1^{(k)}}_S$ cannot be preserved, but only its 
(3, 4, 5) components can be. 
This means that the original field state $\ket{1_\xi}_F$ with $s_1=s_2=0$ 
can be perfectly transferred and stored in the memory subsystem; 
hence the input pulse shape should be synthesized by multiplying the 
classical information $(s_3, s_4, s_5)$ with the basis functions 
$(\nu_3(t), \nu_4(t), \nu_5(t))$, generating as a result 
$\xi(t)=s_3\nu_3(t) + s_4\nu_4(t) + s_5\nu_5(t)$. 
Indeed, in this case, the whole state just after the writing process is 
given by 
\[
    \ket{\Psi(t_1)}
     = \ket{0,0}\otimes
         \Big[s_3 \ket{1,0,0} + s_4 \ket{0,1,0} + s_5 \ket{0,0,1} \Big] 
            \otimes \ket{0}_F, 
\]
and thus the state $s_3 \ket{1,0,0} + s_4 \ket{0,1,0} + s_5 \ket{0,0,1}$ is 
preserved; see Fig.~2~(c).

The idea is the same for the coherent input case. 
That is, the state 
\[
    \ket{\Psi(t_1)}
     = \ket{0,0}\otimes
         \ket{\alpha_3,\alpha_4,\alpha_5}
            \otimes \ket{0}_F 
\]
can be perfectly transferred and stored in the memory subsystem.

\begin{figure}
\centering
\includegraphics[scale=0.4]{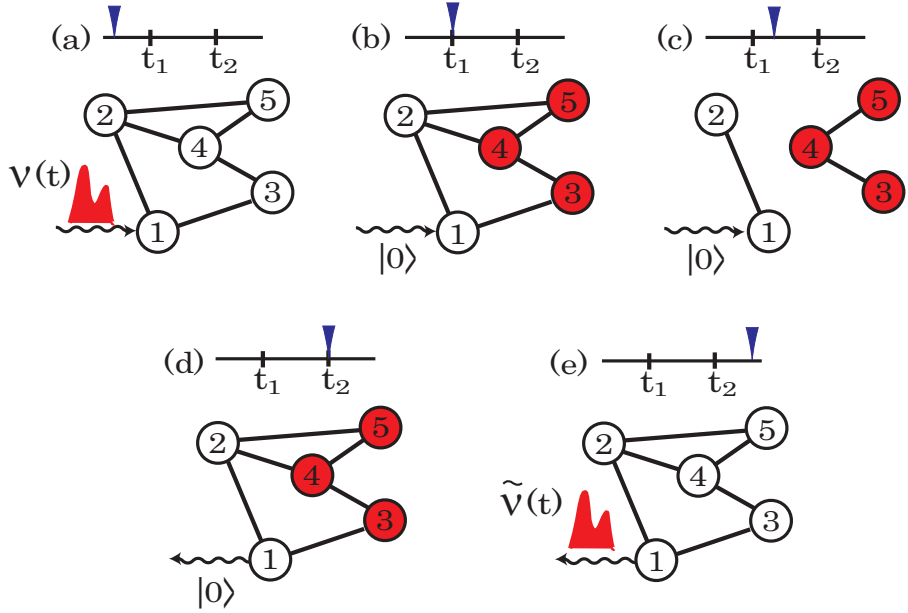}
\caption{
\label{code state transfer}
The perfect memory procedure for the single photon input state in a 5-nodes 
passive linear network. 
The system can be tuned so that the (3, 4, 5) nodes become decoherence free; 
hence these nodes constitute the memory subsystem. 
The (1, 2) nodes does the buffer subsystem. 
(a) The single photon code state with $s_1=s_2=0$ is sent through 
the input optical field with pulse shape $\nu(t)$. 
(b) At time $t=t_1$, the perfect state transfer has been completed. 
The system is then modulated and the memory subsystem is decoupled. 
(c) The transferred state is preserved during the period $[t_1, t_2]$. 
(d) At $t=t_2$ the memory subsystem is again coupled to the buffer subsystem 
and thus the optical field. 
(e) The perfect copy appears in the output optical field with pulse shape 
$\tilde{\nu}(t)$. 
}
\end{figure}

%%%%%%%%%%%%%%%%%%%%%%%%%%%%%%%%%%%%%%%%%%

\subsection{The reading stage}

Suppose that the state has been perfectly stored during the period 
$[t_1, t_2]$; 
hence the reading stage starts at time $t=t_2$ with the initial state 
$\ket{\Psi(t_2)}
= \sum_k s_k \ket{1^{(k)}}_S \otimes \ket{0}_F$ for the 
single photon input case or 
$\ket{\Psi(t_2)}
= \ket{\alpha_1,\ldots,\alpha_n}_S\otimes \ket{0}_F$ for the coherent 
input case; see Fig.~2~(d). 
Note that, as described in Section~5.2, only some elements of $\{s_k\}$ 
or $\{\alpha_k\}$, which represents the classical information of the stored 
state, are not zero. 
To retrieve this initial state, we switch the system matrices so that 
the memory subsystem again couples to the buffer subsystem and thus the 
optical field; 
in particular, we take the same system matrices $\Omega$ and $C$ 
(and thus $A$) as in the writing stage. 
Thus note that $A$ is Hurwitz.

To describe the reading stage, first, we particularly focus on 
the following quantity:
\begin{eqnarray}
& & \hspace*{-5em}
   \int_{t_2}^\infty e^{A^\dagger (t-t_2)} C^\dagger \tilde{b}(t)dt
    = \int_{t_2}^\infty e^{A^\dagger (t-t_2)} C^\dagger 
        \Big( Ca(t)+b(t) \Big)dt
\nonumber \\ & & \hspace*{-3em}
    = \int_{t_2}^\infty e^{A^\dagger (t-t_2)} C^\dagger 
       \Big[ C \Big(
         e^{A(t-t_2)}a(t_2) - e^{A t} \int_{t_2}^t 
            e^{-A s} C^\dagger b(s)ds \Big) + b(t) \Big]dt 
\nonumber \\ & & \hspace*{-3em}
    = \Big[ \int_{t_2}^\infty e^{A^\dagger (t-t_2)} C^\dagger C 
         e^{A (t-t_2)} dt\Big] a(t_2)
     - \int_{t_2}^\infty e^{A^\dagger (t-t_2)} C^\dagger C e^{A t} 
        \Big[ \int_{t_2}^t e^{-A s} C^\dagger b(s)ds \Big] dt
\nonumber \\ & & \hspace*{1em}
    + \int_{t_2}^\infty e^{A^\dagger (t-t_2)} C^\dagger b(t) dt.
\nonumber
\end{eqnarray}
The first term is $a(t_2)$. 
For the second and the third terms, by defining 
$K(t):=\int_{t_2}^t e^{-A s} C^\dagger b(s)ds$, which leads to 
$dK(t)/dt=e^{-At}C^\dagger b(t)$, we find that they become 
\begin{eqnarray}
& & \hspace*{-3em}
    - \int_{t_2}^\infty e^{A^\dagger (t-t_2)} C^\dagger C e^{A t} 
        K(t) dt
    + \int_{t_2}^\infty e^{A^\dagger (t-t_2)} e^{At} \frac{dK(t)}{dt} dt
\nonumber \\ & & \hspace*{1em}
    = e^{-A^\dagger t_2} 
        \int_{t_2}^\infty \frac{d}{dt} 
          \Big[ e^{A^\dagger t} e^{At} K(t) \Big] dt
    = -e^{At_2}K(t_2)=0.
\nonumber
\end{eqnarray}
As a result, we have 
\begin{equation}
\label{observability}
    a^\sharp(t_2) 
      = \int_{t_2}^\infty e^{A^\top (t-t_2)} 
         C^\top \tilde{b}^*(t)dt
      = \int_{-\infty}^\infty \tilde{\nu}(t) \tilde{b}^*(t)dt, 
\end{equation}
where 
\begin{equation}
\label{optimal reading pulse shape vector}
    \tilde{\nu}(t) = e^{A^\top (t-t_2)} C^\top \Theta(t-t_2). 
\end{equation}
As in the previous case, $\tilde{\nu}_i(t)$ are orthonormal; 
$\int_{-\infty}^\infty \tilde{\nu}_i^*(t)\tilde{\nu}_j(t) dt
=\delta_{ij}$. 
Note that $\tilde{\nu}(t)$ is a generalization of a {\it decaying} 
exponential function. 
Moreover, Eq.~\eref{observability} leads to 
\begin{eqnarray}
& & \hspace*{-3em}
    U(t_2,\infty)a^\sharp(t_2)U^*(t_2,\infty)
     = \int_{-\infty}^\infty \tilde{\nu}(t) 
         U(t_2,\infty)\tilde{b}^*(t)U^*(t_2,\infty)dt
\nonumber \\ & & \hspace*{1em}
     = \int_{-\infty}^\infty \tilde{\nu}(t) 
         U(t_2,\infty)U^*(t_2,t)b^*(t)U(t_2,t)U^*(t_2,\infty)dt
\nonumber \\ & & \hspace*{1em}
     = \int_{-\infty}^\infty \tilde{\nu}(t) 
         U(t,\infty)b^*(t)U^*(t,\infty)dt
     = \int_{-\infty}^\infty \tilde{\nu}(t) b^*(t) dt
\nonumber \\ & & \hspace*{1em}
     = [I_S\otimes B^*(\tilde{\nu}_1), \ldots, 
          I_S\otimes B^*(\tilde{\nu}_n)]^\top. 
\nonumber
\end{eqnarray}

{\bf Case I: Single photon state.} 
The initial state is now 
$\ket{\Psi(t_2)}
= \sum_k s_k \ket{1^{(k)}}_S \otimes \ket{0}_F$. 
Then, through the interaction, this state changes to:
\begin{eqnarray}
& & \hspace*{-4em}
    \ket{\Psi(\infty)}
     = U(t_2, \infty)
         \Big[ \sum_k s_k \ket{1^{(k)}}_S \Big] 
            \otimes \ket{0}_F
     = U(t_2, \infty)
         \Big[ \sum_k s_k a_k^*(t_2) \Big] 
            \ket{0,\ldots,0}_S\ket{0}_F
\nonumber \\ & & \hspace*{-0.8em}
     = U(t_2, \infty)
         \Big[ \sum_k s_k a_k^*(t_2) \Big] U^*(t_2, \infty)
            \ket{0,\ldots,0}_S\ket{0}_F
\nonumber \\ & & \hspace*{-0.8em}
     = \sum_k s_k \big[ I_S\otimes B^*(\tilde{\nu}_k) \big]
           \ket{0,\ldots,0}_S\ket{0}_F
     = \ket{0,\ldots,0}_S 
           \otimes \sum_k s_k \ket{1_{\tilde{\nu}_k}}_F. 
\nonumber
\end{eqnarray}
Therefore, certainly the field state recovers the input state 
\eref{single photon input}, which is now carried by the output field with 
pulse shape \eref{optimal reading pulse shape vector}. 
The system state returns to the ground state; see Fig.~2~(d,e).

{\bf Case II: Coherent state.} 
The initial state is 
$\ket{\Psi(t_2)}
= \ket{\alpha_1,\ldots,\alpha_n}_S\otimes \ket{0}_F$. 
Then, through the interaction, this state changes to:
\begin{eqnarray}
& & \hspace*{-4.1em}
    \ket{\Psi(\infty)}
     = U(t_2, \infty)
         \ket{\alpha_1,\ldots,\alpha_n}_S \otimes \ket{0}_F
     = U(t_2, \infty)
         e^{\sum_k \alpha_k a_k^*(t_2) - \alpha_k^* a_k(t_2)}
            \ket{0,\ldots,0}_S\ket{0}_F
\nonumber \\ & & \hspace*{-0.8em}
     = U(t_2, \infty)
         e^{\sum_k \alpha_k a_k^*(t_2) - \alpha_k^* a_k(t_2)} U^*(t_2, \infty)
            \ket{0,\ldots,0}_S\ket{0}_F
\nonumber \\ & & \hspace*{-0.8em}
     = e^{\sum_k \alpha_k B^*(\tilde{\nu}_k) - \alpha_k^* B(\tilde{\nu}_k)}
           \ket{0,\ldots,0}_S\ket{0}_F
     = \ket{0,\ldots,0}_S \otimes \ket{\tilde{f}}_F, 
\nonumber
\end{eqnarray}
where $\ket{\tilde{f}}_F$ is a coherent field state with pulse shape 
\[
      \tilde{f}(t)=\sum_k \alpha_k \tilde{\nu}_k(t). 
\]
Thus, similar to the single photon input case, the stored coherent states 
$\ket{\alpha_1,\ldots,\alpha_n}_S$ leaks into the output field with 
pulse shape \eref{optimal reading pulse shape vector}, and we can retrieve 
the full information about $\{\alpha_k\}$ contained in the coherent field 
state $\ket{\tilde{f}}_F$.

%%%%%%%%%%%%%%%%%%%%%%%%%%%%%%%%%%%%%%%%%%%%
%%%%%%%%%%%%%%%%%%%%%%%%%%%%%%%%%%%%%%%%%%%%
%%%%%%%%%%%%%%%%%%%%%%%%%%%%%%%%%%%%%%%%%%%%

\section{Statistical equations in the writing stage}

Here we derive the time evolution equations of the statistics in the 
writing stage. 
These equations are useful for numerical simulation, as demonstrated 
in the next section.

{\bf Case I: Single photon state.} 
In the case of single photon state, we evaluate the following matrix of 
operators:
\[
    N = a^\sharp a^\top
      = \left[ \begin{array}{c}
           a_1^*  \\
           \vdots \\
           a_n^*  \\
          \end{array} \right]
       [a_1, \ldots, a_n]. 
\]
The photon is distributed in the system according to the statistics represented 
by the correlation matrix 
$\mean{N}_{11}=(\bra{0,1_\xi}a^*_i a_j\ket{0,1_\xi})$. 
The time-evolution equation of $\mean{N}_{11}$ is, together with 
the vector $\mean{a^\sharp}_{10}=[\bra{0,1_\xi}a^*_1\ket{0,0}, \ldots, 
\bra{0,1_\xi}a^*_n\ket{0,0}]^\top$, given by 
\begin{eqnarray}
\label{photon dis eq1}
& & \hspace*{0em}
    \frac{d}{dt}\mean{N}_{11} 
     = A^\sharp\mean{N}_{11} + \mean{N}_{11}A^\top
        - \xi^*(t) C^\top \mean{a^\sharp}_{10}^\dagger
        - \xi(t) \mean{a^\sharp}_{10} C^\sharp, 
\\ & & \hspace*{0em}
\label{photon dis eq2}
    \frac{d}{dt}\mean{a^\sharp}_{10} 
     = A^\sharp \mean{a^\sharp}_{10}
        - C^\top\xi^*(t). 
\end{eqnarray}
The solution of Eq.~\eref{photon dis eq2} is readily obtained as 
\[
    \mean{a^\sharp(t)}_{10}
     = - e^{A^\sharp t}
           \int_{t_0}^t e^{-A^\sharp s} C^\top \xi^*(s) ds, 
\]
where $\mean{a^\sharp(t_0)}_{10}=0$ is used. 
In general, the Lyapunov differential equation 
$dQ/dt=AQ+QA\dgg+R$, with $R(t)=R^\dagger(t)$ time varying, has the 
solution of the form
\[
    Q(t) = e^{A(t-t_0)}Q(t_0)e^{A\dgg(t-t_0)}
        + e^{At}\Big( 
            \int_{t_0}^t e^{-As}R(s)e^{-A\dgg s} ds
              \Big)e^{A\dgg t}. 
\]
If $A$ is Hurwitz, in the limit of $t_0\rightarrow -\infty$, this becomes 
\[
    Q(t) = e^{At}\Big( 
            \int_{-\infty}^t e^{-As}R(s)e^{-A\dgg s} ds
              \Big) e^{A\dgg t}. 
\]
Using this result and the expression \eref{optimal pulse shape vector}, 
we have 
\[
    \mean{N(t_1)}_{11}
     = \Big( \int_{-\infty}^\infty \xi(t) \nu(t)\dgg dt\Big)\dgg
         \Big( \int_{-\infty}^\infty \xi(t) \nu(t)\dgg dt\Big).
\]
If we send the single photon code state over the input field with 
rising exponential pulse shape $\xi(t)=\sum_k s_k \nu_k(t)$, then 
we have $\mean{N(t_1)}_{11}=(s_i^* s_j)$ due to 
$\int_{-\infty}^\infty \nu_i^*(t)\nu_j(t) dt=\delta_{ij}$. 
This means that the input single photon state is distributed among 
the network so that the $k$th node has the mean photon number 
$|s_k|^2$ at time $t_1$.

{\bf Case II: Coherent state.} 
In this case the statistics is more convenient, because a coherent state 
is completely charactered only by its mean and variance. 
In particular, the dynamics of the mean, $m(t)=\mean{a(t)}$, was already 
obtained in Eq.~\eref{dynamics of m}, with $f(t)$ particularly given by 
$f(t) = - Ce^{-A^\dagger (t-t_1)}\alpha \Theta(t_1-t)$ in Eq.~\eref{f in section 5}. 
Hence it is immediate to obtain the solution in $t\leq t_1$: 
\begin{eqnarray}
& & \hspace*{0em}
     m(t) = e^{A(t-t_0)}m(t_0) - e^{At} \int_{t_0}^t e^{-As} C^\dagger f(s)ds
\nonumber \\ & & \hspace*{2em}
     = e^{A(t-t_0)}m(t_0) 
                  + e^{At} \Big(\int_{t_0}^t \frac{d}{ds}(e^{-As} e^{-A^\dagger s}) ds \Big)
                       e^{A^\dagger t_1}\alpha
\nonumber \\ & & \hspace*{2em}
     = e^{A(t-t_0)}m(t_0) 
           + e^{-A^\dagger(t-t_1)}\alpha
               - e^{A(t-t_0)}e^{A^\dagger(t_1-t_0)}\alpha.
\nonumber 
\end{eqnarray}
Hence, by taking the limit $t_0\rightarrow -\infty$, we have 
$m(t_1)=\alpha$; i.e. $\mean{a_k(t_1)}=\alpha_k$. 
Also, we evaluate the covariance matrix 
$V=\mean{\Delta a^\sharp \Delta a^\top}$ with $\Delta a = a -\mean{a}$, 
which takes zero if and only if the state is a coherent state. 
Similar to the single photon case, we find that $V(t)$ obeys 
$dV(t)/dt = A^\sharp V(t) + V(t)A^\top$, which readily yields 
$V(t)=e^{A^\sharp(t-t_0)} V(t_0) e^{A^\top(t-t_0)}\rightarrow O$ as 
$t_0\rightarrow -\infty$. 
As a result, the $k$th node becomes the coherent state $\ket{\alpha_k}$ 
at time $t=t_1$. 
We note that the mean of the output field is $\tilde{f}(t)=Cm(t)+f(t)=0$ for 
all $t\leq t_1$; thus the zero-dynamics principle is certainly satisfied.

%%%%%%%%%%%%%%%%%%%%%%%%%%%%%%%%%%%%%%%%%%%%
%%%%%%%%%%%%%%%%%%%%%%%%%%%%%%%%%%%%%%%%%%%%
%%%%%%%%%%%%%%%%%%%%%%%%%%%%%%%%%%%%%%%%%%%%

\section{Example: Perfect memory network with atomic ensembles}

This section is devoted to study a passive linear network composed of 
atomic ensembles, which contains a tunable DF component. 
A numerical simulation will demonstrate how the input field state is 
transferred to the memory subsystem and how the input pulse shape to be 
engineered for perfect memory looks like.

%%%%%%%%%%%%%%%%%%%%%%%%%%%%%%%%%%%%%%%%%%%%

\subsection{The atomic ensembles trapped in a cavity}

\begin{figure}
\centering
\includegraphics[scale=0.53]{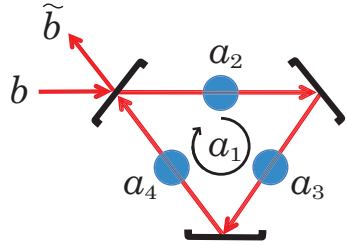}
\caption{
\label{atom example}
The passive linear network composed of three large atomic ensembles and  
a ring cavity. 
$a_1$ denotes the cavity mode, and $a_k~(k=2,3,4)$ is the annihilation 
operator approximating the collective lowering operator of the $k$th 
atomic ensemble. 
}
\end{figure}

The system is three large atomic ensembles trapped in a single-mode 
cavity, depicted in Fig.~\ref{atom example}; 
a detailed description of this system is found in e.g. 
\cite{Duan,Kuzmich 2006,Parkins2006,Parkins2007,
Ficek2009,PolzikRMP2010}. 
The annihilation operator $a_1$ represents the cavity mode, and 
$a_k~(k=2,3,4)$ is the annihilation operator approximating the 
collective lowering operator of the $k$th ensemble. 
The internal cavity light field and the $k$th ensemble interact with each 
other through external pulse lasers with Rabi frequencies 
$\omega_k$ and $\omega'_k$. 
The coupling Hamiltonian is given by 
\begin{eqnarray} 
    H_{\rm ac}
       =\frac{\sqrt{N}\mu}{2\delta}
           \sum_{k=2}^4 \Big[ a_1^* (\omega_k e^{i\phi_k} a_k 
                                                      + \omega'_k e^{i\phi'_k}a_k^*) + \mathrm{H.c.} 
           \Big], 
\end{eqnarray}
where $\phi_k \in [0,2\pi)$ is the laser phase, $N$ is the number of atoms 
in each ensemble, $\mu$ is the coupling strength, and $\delta$ is the 
detuning. 
The spontaneous emission of each atom is negligible for typical atoms 
such as $^{87}$Rb. 
We also assume that the second and third ensembles can be manipulated 
via external magnetic fields, which introduce the self Hamiltonian 
$H_{\rm a} = \Delta a_2^*a_2 - \Delta a_3^*a_3$ with $\Delta$ denoting 
the tunable strength of the magnetic field. 
We here set the parameters as $\omega_k=\omega>0,~\omega'_k=0$ 
and $\phi_k=\pi/2$ for $k=2,3,4$, and define 
$g=\sqrt{N}\mu \omega/2\delta$; 
then, the total system Hamiltonian is given by 
\begin{eqnarray}
& & \hspace*{-2em}
    H = H_{\rm a} + H_{\rm ac} 
        = \Delta a_2^*a_2 - \Delta a_3^*a_3 
         + ig a_1^*(a_2+a_3+a_4) - ig (a_2^*+a_3^*+a_4^*)a_1
\nonumber \\ & & \hspace*{-1.1em}
       = [a_1^*, a_2^*, a_3^*, a_4^*]
           \left[ \begin{array}{cccc}
               0 & ig & ig & ig  \\
               -ig & \Delta & 0 & 0  \\
               -ig & 0 & -\Delta & 0  \\
               -ig & 0 & 0 & 0  \\
           \end{array} \right]
           \left[ \begin{array}{c}
               a_1 \\
               a_2 \\
               a_3 \\
               a_4 \\
           \end{array} \right]
        = a^\dagger \Omega a.
\nonumber
\end{eqnarray}
The cavity field couples to an external optical field with continuous mode 
$b(t)$ used for state transfer, at the beam splitter with transmissivity 
proportional to $\kappa$; 
this means that the system-field coupling Hamiltonian 
$H_{\rm int}(t) = i[b^*(t)Ca - a^\dagger C^\dagger b(t)]$, which was 
defined above Eq.~\eref{general unitary}, is specified with 
$Ca=\sqrt{\kappa}a_1$. 
Consequently, the system matrices are given by 
\[
     A = -i\Omega - \frac{1}{2}C^\dagger C 
        = \left[ \begin{array}{cccc}
           -\kappa/2 & g & g & g  \\
           -g & -i\Delta & 0 & 0  \\
           -g & 0 & i\Delta & 0  \\
           -g & 0 & 0 & 0  \\
          \end{array} \right],~~~
    C = [\sqrt{\kappa}, 0, 0, 0]. 
\]
Note that this passive linear system can be physically realized in some 
other systems, such as a mechanical oscillator array connected in 
a single mode cavity.

%%%%%%%%%%%%%%%%%%%%%%%%%%%%%%%%%%%%%%%%%%%%

\subsection{The perfect memory procedure}

We can prove that, when $\Delta\neq 0$, the matrix $A$ is Hurwitz; 
i.e. the real part of all the eigenvalues of $A$ is negative. 
A convenient way to see this fact is to use the property that the controllability 
matrix $[C^\dagger, AC^\dagger, \ldots, A^{n-1}C^\dagger]$ is of full rank 
iff $A$ is Hurwitz \cite{Guta Yamamoto}. 
Thus the system does not contain a DF component when $\Delta\neq 0$. 
On the other hand, if we turn off the magnetic field and set $\Delta=0$, then 
a DF subsystem appears, as shown below. 
Let us take the following unitary matrix: 
\[
    U = \left[ \begin{array}{cccc}
           1 & 0 & 0 & 0  \\
           0 & 1/\sqrt{3} & 2/\sqrt{6} & 0  \\
           0 & 1/\sqrt{3} & -1/\sqrt{6} & 1/\sqrt{2}  \\
           0 & 1/\sqrt{3} & -1/\sqrt{6} & -1/\sqrt{2}  \\
          \end{array} \right].
\]
This transforms the system equation to
\[
    \dot{a}'(t)=A'a'(t) - C'\mbox{}\dgg b(t),~~~
    \tilde{b}(t)=C'a'(t)+b(t), 
\]
where 
\begin{eqnarray}
\label{transformed matrices}
& & \hspace*{0em}
    a' = U^\dagger a 
       = \left[ \begin{array}{c}
           a_1 \\
           (a_2+a_3+a_4)/\sqrt{3} \\
           (2a_2-a_3-a_4)/\sqrt{6} \\
           (a_3-a_4)/\sqrt{2} \\
          \end{array} \right],
\nonumber \\ & & \hspace*{0em}
    A'=U\dgg AU 
       = \left[ \begin{array}{cccc}
            -\kappa/2 & \sqrt{3}g & 0 & 0  \\
            -\sqrt{3}g & 0 & -\sqrt{2}i\Delta/2 & \sqrt{6}i\Delta/6  \\
            0 & -\sqrt{2}i\Delta/2 & -i\Delta/2 & -\sqrt{3}i\Delta/6  \\
            0 & \sqrt{6}i\Delta/6 & -\sqrt{3}i\Delta/6 & i\Delta/2  \\
         \end{array} \right],
\nonumber \\ & & \hspace*{0em}
     C' = CU 
       = [\sqrt{\kappa},~0,~0,~0].
\end{eqnarray}
Therefore, when $\Delta=0$, the system takes a form of 
Eq.~\eref{general DF system}. 
That is, $a_{\rm M}=[a'_3, a'_4]^\top$ is not affected by the incoming field 
$b(t)$ and it does not appear in the output field $\tilde{b}(t)$; 
hence $a_{\rm M}=[a'_3, a'_4]^\top$ is the memory subsystem that can be 
switched to a DF or non-DF subsystem, just by controlling the external 
magnetic field. 
This two-mode subsystem works as a perfect memory that preserves any 
state of the form $s_3\ket{0,0,1,0}+s_4\ket{0,0,0,1}$ in the case of single 
photon state or $\ket{0,0,\alpha_3, \alpha_4}$ in the case of coherent state. 
Note that $a_3'$ and $a_4'$ depend on the atomic modes $(a_2, a_3, a_4)$ 
and not on the cavity mode $a_1$, implying that the state is indeed 
stored in the atomic ensembles. 
Also it should be remarked that $a_3'$ and $a_4'$ take the form of 
continuous-variable {\it syndromes} used for quantum error correction 
\cite{Braunstein 1998,Lloyd 1998}.

\begin{figure}
\centering
\includegraphics[scale=0.4]{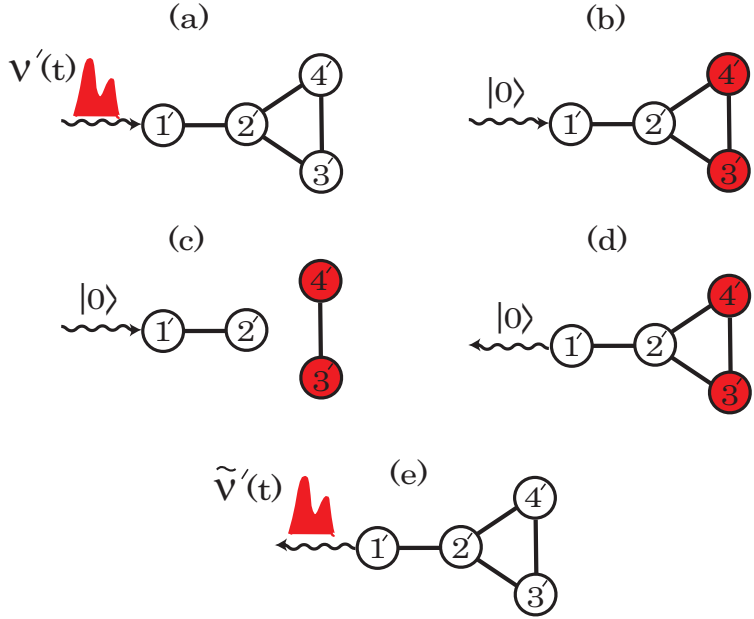}
\caption{
\label{fig example}
The memory procedure for the passive linear network composed of 
three atomic ensembles trapped in a single-mode cavity. 
The number $k'$ indicates the subsystem with mode $a_k'$. 
(a) The single photon state is sent through the input optical field with pulse 
shape $\nu'(t)$, where in this stage the magnetic field is turned on 
($\Delta\neq 0$). 
(b) At time $t=t_1$ the system acquires the state 
$s_3\ket{0,0,1,0}+s_4\ket{0,0,0,1}$; 
that is, the input state is perfectly transferred into the 3rd and 4th nodes. 
(c) Then the magnetic field is turned off ($\Delta=0$) so that the memory 
subsystem with modes $(a_3', a_4')$ is decoupled from the buffer 
subsystem with modes $(a_1', a_2')$ and the input-output optical field; 
hence it becomes decoherence free and the transferred state is perfectly 
preserved. 
(d) At $t=t_2$ we again set $\Delta\neq 0$. 
The memory subsystem again couples to the buffer subsystem and the 
optical field. 
(e) The perfect copy appears in the output field with the pulse shape 
$\tilde{\nu}'(t)$. 
}
\end{figure}

Here we describe the concrete procedure of the writing, storage, and reading 
processes, in the case of single photon input; 
see Fig.~\ref{fig example}. 
\begin{itemize}

\item 
A single photon field state is prepared in the form 
$s_3\ket{1_{\nu'_3}}+s_4\ket{1_{\nu'_4}}$, where $\nu'_3(t)$ and $\nu'_4(t)$ 
are the third and fourth elements of the vector of rising exponential functions 
$\nu'(t) = - e^{-A'\mbox{}^\sharp (t-t_1)} C'\mbox{}^\top \Theta(t_1-t)$ 
with $A'$ and $C'$ given in Eq.~\eref{transformed matrices}. 
Note in this stage the magnetic field is ON; $\Delta\neq 0$. 

\item
The field couples to the system until $t\leq t_1$. 
The perfect state transfer is achieved in the end, at $t=t_1$, by sending 
the input state over the optical field with pulse shape $\nu'(t)$ described 
above. 
The whole state changes to 
$\ket{0,0}\otimes(s_3\ket{1,0}+s_4\ket{0,1})\otimes\ket{0}_F$. 

\item
We turn off the magnetic field and set $\Delta=0$; 
then the memory subsystem with modes $(a'_3, a'_4)$ becomes decoherence 
free and its state $s_3\ket{1,0}+s_4\ket{0,1}$ is preserved during an arbitrary 
time interval $[t_1, t_2]$. 

\item
At a later time $t_2$, we turn on the magnetic field  (i.e. set 
$\Delta\neq 0$) to retrieve the stored state. 
Then the memory subsystem again couples to the optical field, and the 
perfect copy $s_3\ket{1_{\tilde{\nu}'_3}}+s_4\ket{1_{\tilde{\nu}'_4}}$ 
appears in the output field with the pulse shape specified by 
$\tilde{\nu}'(t) = e^{A'\mbox{}^\top (t-t_2)} C'\mbox{}^\top \Theta(t-t_2)$. 

\end{itemize}

Recall that the optimal input pulse shape is determined by the 
properties (zeros) of the memory system and this corresponds to the 
impedance matching mentioned in Section 3.2. 
Now we should note that an additional matching condition is not imposed 
on the interaction between the cavity mode and the atomic ensembles, 
although perfect state transfer from the former to the latter is certainly 
achieved. 
This result seems to be inconsistent with the fact obtained in 
\cite{Afzelius 2010,Moiseev PRA 2010,Moiseev PRA 2013,Chaneliere 2014}, 
showing that perfect state transfer from a cavity to an inhomogeneously 
broadened (IB) atomic ensemble requires a strict impedance matching 
between them. 
But there is a clear difference between our case and those studies; 
in the case dealing with the IB ensemble, due to the matching condition, 
an input field state with {\it arbitrary} (yet within a finite band-width) 
temporal shape is allowed to be completely absorbed into the ensemble 
(see e.g. \cite{Special issue,Tittel 2010} for the recent experimental results), 
while in our case the optimal pulse shape has to be strictly specified. 
Exploring a combined schematic of these two memory procedures, which 
would allow weaker pulse shaping and weaker impedance matching, 
should be an interesting future work.

%%%%%%%%%%%%%%%%%%%%%%%%%%%%%%%%%%%%%%%%%%%%

\subsection{Numerical simulation}

\begin{figure}
\centering
\includegraphics[scale=0.52]{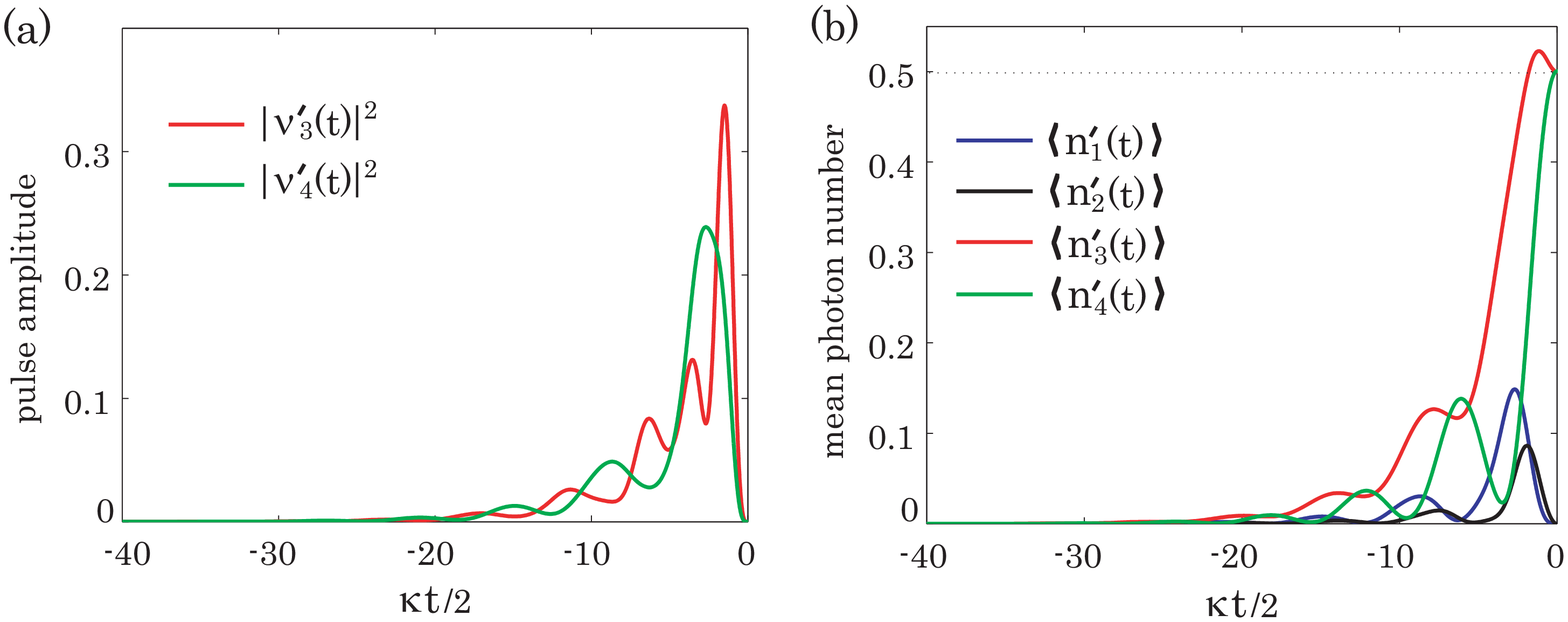}
\caption{
\label{network simulation}
(a) Time evolutions of the absolute value of $\nu'_3(t)$ (red) and $\nu'_4(t)$ 
(green). 
(b) Time evolution of the mean photon number at the $i$th nodes, 
$\mean{n'_i(t)}=\mean{a_i'\mbox{}^*(t)a'_i(t)}$. 
The blue, black, red, and green lines represent the time evolutions 
of $\mean{n_1'(t)}, \mean{n_2'(t)}, \mean{n_3'(t)}$, and $\mean{n_4'(t)}$, 
respectively. 
}
\end{figure}

Here we demonstrate a numerical simulation of the writing stage of 
the above memory procedure. 
The parameters are set to $\kappa=2$, $g=1$, and $\Delta=1$; 
note again in this stage $\Delta\neq 0$ and there is no DF subsystem. 
The input is a single photon field state with coefficients $s_3=s_4=1/\sqrt{2}$, 
which is carried by the optical field with pulse shape $\nu'_3(t)$ and 
$\nu_4'(t)$ as mentioned above. 
The initial time is $\kappa t_0/2=-40$ and the stopping time is $t_1=0$.

First, Fig.~\ref{network simulation}~(a) shows the absolute value of 
$\nu'_3(t)$ and $\nu_4'(t)$. 
These are the pulse shapes we need to correctly engineer for the 
desirable perfect state transfer. 
A notable point is that they are not anymore of a rising exponential shape 
such as Eq.~\eref{RE example}; 
particularly they take the value zero at the stopping time $t_1=0$. 
A similar non-rising exponential pulse function was also found in 
\cite{Moiseev JPB 2012}, achieving perfect state transfer in an 
integrated quantum memory system. 
It looks that we can realize this kind of pulse shape by combining some 
Gaussian wave packets, which might be a desirable feature from the 
engineering viewpoint.

Next, Fig.~\ref{network simulation}~(b) shows the time-evolutions of 
the mean photon number of each node, i.e. 
$\mean{n_i'(t)}=\mean{a_i'\mbox{}^*(t)a_i'(t)}~(i=1,2,3,4)$, which can be 
computed by numerically solving Eqs.~\eref{photon dis eq1} and 
\eref{photon dis eq2}. 
As expected from the theory, the memory subsystem with modes 
$(a_3', a_4')$ perfectly acquires the photon with mean photon number 
$\mean{n_3'(0)}=\mean{n_4'(0)}=0.5$ at $t_1=0$. 
We should note that the transportation of the photon from the input field 
to the memory subsystem occurs rapidly only in the last few period; 
in fact, almost all the energy contained in the input pulses $\nu'_3(t)$ and 
$\nu'_4(t)$ is confined in this short period. 
Hence, we need to be very careful to stop the writing process at the accurate 
time $t_1=0$, because the desired state 
$\ket{0,0}\otimes(\ket{1,0}+\ket{0,1})/\sqrt{2}$ is fragile in the following 
sense. 
For instance if we turn off the magnetic field a bit earlier than $t_1=0$, 
say $t_1=\kappa t_1/2=-1$, then the whole system's state generated is 
roughly 
$0.1\ket{1,0,0,0}+0.1\ket{0,1,0,0}+0.52\ket{0,0,1,0}+0.4\ket{0,0,0,1}$ 
(unnormalized); 
thus the state of the memory subsystem becomes a mixed state (unnormalized)
\[
     \rho_{3'4'} = 0.01 \ket{0,0}\bra{0,0}
       + \Big( 0.52\ket{1,0} + 0.4\ket{0,1}\Big)\Big( 0.52\bra{1,0} + 0.4\bra{0,1}\Big)
\]
due to the decoherence added to the buffer subsystem with modes 
$(a_1', a_2')$ during the storage period. 
Hence, an important future work is to find a suitable set of parameters 
$(\kappa, g, \Delta)$ so that the time-evolutions of the mean photon numbers 
of the memory subsystem become as flat as possible at the stopping time $t_1$.

%%%%%%%%%%%%%%%%%%%%%%%%%%%%%%%%%%%%%%%%%%%%
%%%%%%%%%%%%%%%%%%%%%%%%%%%%%%%%%%%%%%%%%%%%
%%%%%%%%%%%%%%%%%%%%%%%%%%%%%%%%%%%%%%%%%%%%

\section{Conclusion}

In this paper, for a general passive linear system, we have provided 
a designing method of input pulse shape that perfectly transports a 
single photon or coherent field state to a memory subsystem, which 
can be switched to a DF subsystem. 
The method is general and simple, so it can be directly applied to 
a large-scale network; 
in fact, in the example studied in Section~7, we found that the explicit form 
of $\nu'_3(t)$ and $\nu'_4(t)$ are readily obtained. 
The results are based on the zero-dynamics principle. 
Although in this paper this principle was used only for synthesizing the 
input pulse shape, it is indeed a wide concept that works in a more general 
situation. 
For example, the zero-dynamics principle can be applied to the case 
where, instead of pulse shaping of the input field, some time-varying 
controllable parameters of the system should be engineered due to practical 
limitation; 
also the system can be nonlinear; 
further, we could deal with an inhomogeneously broadened atomic ensemble 
memory that allows an arbitrary temporal shape for perfect state transfer, 
which was discussed in Section~7.2. 
In any case, following the zero-dynamics principle, we should design 
the system so that the output is zero or more generally the output is 
minimized. 
Moreover, the zero-dynamics corresponds to the time-evolution of a state 
free from any energy loss, thus it represents a coherent, yet non-unitary, 
gate operation on the system state for quantum information processing; 
that is, designing a desired manipulation of a state in an open system is 
no more than designing a desired zero-dynamics. 
All these problems will be addressed in future works.

%%%%%%%%%%%%%%%%%%%%%%%%%%%%%%%%%%%%%%%%%%%%
%%%%%%%%%%%%%%%%%%%%%%%%%%%%%%%%%%%%%%%%%%%%
%%%%%%%%%%%%%%%%%%%%%%%%%%%%%%%%%%%%%%%%%%%%

\ack
The authors acknowledges helpful discussions with M. R. Hush and 
A. R. R. Carvalho. 
This research was supported by the Australian Research Council Centre 
of Excellence for Quantum Computation and Communication Technology 
(project number CE110001027), and AFOSR Grant FA2386-12-1-4075. 
Also NY's work was supported by JSPS grant-in-aid number 40513289.

%%%%%%%%%%%%%%%%%%%%%%%%%%%%%%%%%%%%%%%%%%%%
%%%%%%%%%%%%%%%%%%%%%%%%%%%%%%%%%%%%%%%%%%%%
%%%%%%%%%%%%%%%%%%%%%%%%%%%%%%%%%%%%%%%%%%%%

\appendix

\section{Active memory system}

In this paper, we thoroughly study a passive linear system, but there are 
many systems containing an {\it active} component. 
In general, for such an active system the energy balance identity 
\eref{balanced equality} does not hold, hence the zero-dynamics principle 
does not anymore mean the perfect energy transfer. 
Hence, it should be worth doing a case study to see if an active system could 
allow perfect state transfer.

Let us consider the following active system: 
\[
     \frac{d}{dt}\left[ \begin{array}{c}
           a  \\
           a^*  \\
          \end{array} \right]
     =-\frac{1}{2}\left[ \begin{array}{cc}
           \kappa & -\epsilon  \\
           -\epsilon & \kappa  \\
          \end{array} \right]
       \left[ \begin{array}{c}
           a  \\
           a^*  \\
          \end{array} \right]
      -\sqrt{\kappa}
         \left[ \begin{array}{c}
           b  \\
           b^*  \\
          \end{array} \right]. 
\]
In optics, this represents the dynamics of an optical parametric oscillator, 
where $\epsilon$ denotes the squeezing strength \cite{GardinerBook,
WisemanBook}. 
Note that the system becomes passive if $\epsilon=0$. 
The above equation can be explicitly solved: 
\begin{eqnarray}
& & \hspace*{-4em}
    a^*(t_1)
      = e^{-\kappa(t_1-t_0)/2}
         \Big[ a(t_0) \sinh(\epsilon(t_1-t_0)/2)
             + a^*(t_0) \cosh(\epsilon(t_1-t_0)/2) \Big] 
\nonumber \\ & & \hspace*{-3em}
    \mbox{}
     - \sqrt{\kappa}
             \int_{t_0}^{t_1}
         e^{-\kappa(t_1-s)/2}
          \Big[ \sinh(\epsilon(t_1-s)/2) b(s)ds
              + \cosh(\epsilon(t_1-t_0)/2) b^*(s)ds \Big]. 
\nonumber
\end{eqnarray}
Unlike the passive case, the field annihilation operator $b^*(t)$ 
appears in the equation. 
Then under the same setting taken in Section~3 where the input field state 
is given by a superposition of the vacuum and $\ket{1_{\xi_1}}_F$ with the 
pulse shape function $\xi_1(t)$ given below, we obtain ($t_0\rightarrow -\infty$ 
and $t_1=0$)
\begin{eqnarray}
& & \hspace*{-5.6em}
    \ket{\Psi(t_1)}
     = U(t_0, t_1) \ket{0}_S (\alpha \ket{0}_F + \beta \ket{1_{\xi_1}}_F)
\nonumber \\ & & \hspace*{-5em}
     = \Big[ \alpha \ket{0}_S 
          + \beta \sqrt{
                \frac{2(\kappa^2-\epsilon^2)}{2\kappa^2-\epsilon^2} 
                   } \ket{1}_S
             \Big] \otimes \ket{0}_F
%\nonumber \\ & & \hspace*{6em}
%    \mbox{}
     - \frac{\beta \epsilon}{\sqrt{2\kappa^2-\epsilon^2}}
          U(t_0, t_1) B(\xi_2) U^*(t_0, t_1) \ket{0}_S\ket{0}_F, 
\nonumber
\end{eqnarray}
where 
\[
\hspace{-4em}
     \xi_1(t)=-\sqrt{\frac{2\kappa(\kappa^2-\epsilon^2)}{2\kappa^2-\epsilon^2}}
        e^{\kappa t/2}\cosh(\epsilon t/2),~~~
     \xi_2(t)=\sqrt{\frac{2\kappa(\kappa^2-\epsilon^2)}{\epsilon^2}}
        e^{\kappa t/2}\sinh(\epsilon t/2).
\]
This equation implies that, when $\epsilon\neq 0$, the perfect state transfer 
is impossible due to the third term, which clearly stems from the active 
element of the system. 
To carry out efficient state transfer, we need some approximation; 
in the above case, if $\kappa$ is much bigger than $\epsilon$, then the 
system state becomes approximately the desired one to be stored. 
Another example is found in \cite{Muschik2006}, where the system 
is an atomic ensemble containing an active component, but by introducing 
a fast oscillating magnetic field it is approximated by a passive one, which 
was further shown to be a perfect memory.

%%%%%%%%%%%%%%%%%%%%%%%%%%%%%%%%%%%%%%%%%%%%

\section{Dark state principle}

The basic idea of dark state principle is as follows. 
For a system coupled to a probe field, we continuously monitor the system 
by a photo detector measuring the output field; 
then if the detector counts no photon, this means that the system is in 
a {\it dark state} and has a time evolution without loss of energy. 
Here we apply this dark state principle to the writing problem discussed 
in Section~3 and derive the same result; 
that is, in this sense, the zero-dynamics principle and the dark state 
principle are equivalent, though there is a big difference in practice 
as shown below.

First let us consider the case where we want to send a coherent field state 
to the system. 
In general, if we use a photon counter to estimate the system observables, 
our state (knowledge) conditioned on the measurement results is updated 
by the following stochastic master equation \cite{Bouten2007,Gough 2012} 
(the scattering operator is now set to be the identity): 
\begin{equation*}
\hspace{-2.5cm}
   d\rho = \big({\cal L}\rho + [\rho, L^*]\alpha + [L, \rho]\alpha^*\big)dt 
      + \Big[ \frac{1}{\cal N}
                      (L\rho L^* + \alpha^*L\rho + \alpha \rho L^* + |\alpha|^2\rho)
                    - \rho \Big]
               (dY - {\cal N}dt), 
\end{equation*}
where 
\[
\hspace*{-5em}
    {\cal L}\rho = -i[H, \rho] 
                   + L\rho L^* - L^* L\rho/2 - \rho L^* L/2,~~ 
    {\cal N} = \Tr\Big[ \rho\big(L^*L + \alpha^*L 
           + \alpha L^* + |\alpha|^2\big) \Big].
\]
$\alpha(t)$ is the pulse shape of the input coherent light field and $dY(t)$ 
is the measurement result (0 or 1) obtained during the small time interval 
$[t, t+dt)$. 
Also $H$ and $L$ are the system operators. 
Now since the ensemble averaging over the measurement results leads to 
a standard master equation, we have ${\mathbb E}(dY - {\cal N}dt)=0$. 
Then, the counting probability of the measurement result ``1" during 
$[t, t+dt)$ is given by ${\mathbb P}_1(dt)={\mathbb E}(dY)={\cal N}dt$. 
Hence if ${\cal N}=0~\forall t$, the system is in a dark state and loses no energy 
into the output field; 
the state satisfying this condition is called the dark state. 
In our case where the system is the single-mode passive linear system 
with $H=0$ and $L=\sqrt{\kappa}a$, the dark state can be specified 
to a coherent state $\rho(t)=\ket{\beta(t)}\bra{\beta(t)}$ because we 
now know that the system's state is always a coherent state. 
The condition ${\cal N}=0$ then becomes 
$\kappa|\beta|^2 + \sqrt{\kappa}(\alpha\beta^* + \alpha^* \beta) 
+ |\alpha|^2 = 0$, which yields $\beta=-\alpha/\sqrt{\kappa}$. 
Now, under the dark state condition the time evolution of the conditional 
state is identical to that of the averaged one, which consequently leads to 
$\dot{\beta}=-\kappa \beta/2 - \sqrt{\kappa}\alpha$. 
These two equations yield $\dot{\alpha}=\kappa \alpha/2$, thus 
the input pulse shape must be a rising exponential function 
$\alpha(t)=e^{\kappa(t-t_0)/2}\alpha_0$.

Next let us consider the case where the input is a single photon field state. 
As in the above case, the dark state principle is represented in terms of 
the conditional state subjected to the single-photon stochastic master 
equation; 
see Eq.~(43) in \cite{Gough 2012}. 
In this case the probability to obtain the measurement result ``1" during 
$[t, t+dt)$ is given by 
\[
\hspace{-1.4cm}
     {\mathbb P}_1(dt)={\cal N}dt,~~~
     {\cal N} = \Tr(\rho^{11}L^*L) + \Tr(\rho^{10}L)\xi^* 
                      + \Tr(\rho^{01}L^*)\xi + \Tr(\rho^{00}I)|\xi|^2, 
\]
where $\xi(t)$ is the temporal pulse shape of the single photon field. 
$\rho^{ij}(t)$ are the operators characterizing the conditional state. 
Under the dark state condition ${\cal N}=0$, they obey 
\[
\hspace{-1cm}
    \dot{\rho}^{11} = {\cal L}\rho^{11} 
                 + [\rho^{01}, L^*]\xi + [L, \rho^{10}]\xi^*,~~
    \dot{\rho}^{10} = {\cal L}\rho^{10} + [\rho^{00}, L^*]\xi,~~
    \dot{\rho}^{00} = {\cal L}\rho^{00},
\]
and $\rho^{01}=(\rho^{10})^*$, which are identical to the single-photon 
master equation \cite{Zoller 1998}. 
The initial conditions are $\rho^{11}(0)=\rho^{00}(0)=\ket{0}\bra{0}$ and 
$\rho^{10}(0)=\rho^{01}(0)=0$. 
In our case $H=0$ and $L=\sqrt{\kappa}a$, these differential equations 
can be explicitly solved, yielding 
$\rho^{11}=(1-x)\ket{0}\bra{0}+x\ket{1}\bra{1}$, 
$\rho^{01}=z\ket{0}\bra{1}$, and $\rho^{00}=\ket{0}\bra{0}$, where 
$x(t)$ and $z(t)$ satisfy $\dot{x}=-\kappa x-\sqrt{\kappa}(\xi z + \xi^* z^*)$ 
and $\dot{z}=-\kappa z/2-\sqrt{\kappa}\xi^*$. 
Substituting these solutions $\rho^{ij}(t)$ for the dark state condition 
${\cal N}=0$, we have $\kappa x+\sqrt{\kappa}(\xi z + \xi^* z^*)+|\xi|^2=0$. 
Combining these three equations, we end up with the relation 
$\dot{\xi}=\kappa \xi/2$ and thus see that the input pulse shape has 
to be a rising exponential function $\xi(t)=e^{\kappa(t-t_0)/2}\xi_0$.

Summarizing, we have recovered the same result obtained in Section~3, 
showing that the dark state principle is equivalent to the zero-dynamics 
principle. 
Both principles require no energy leaking from the system into the output 
field, but their approaches are different; 
the dark state principle is represented in the Schr\"{o}dinger picture, 
while the Heisenberg picture is used to describe the zero-dynamics 
principle. 
As a result, in the former case we need to solve the master equation, which 
is sometimes a hard task as demonstrated above especially in the single 
photon field case. 
On the other hand, we have seen in Section 4.2 that the zero-dynamics 
principle allows us to derive the rising exponential function very easily, 
even in the general setup; 
also the transfer-function-based treatment of the principle is notable 
and possibly very useful from the viewpoint of the applicability of the 
linear response theory to quantum memory. 
Of course these special advantages appear particularly in the linear case, 
and for more general nonlinear memory systems we should be careful in 
choosing the approach.

%%%%%%%%%%%%%%%%%%%%%%%%%%%%%%%%%%%%%%%%%%%%
%%%%%%%%%%%%%%%%%%%%%%%%%%%%%%%%%%%%%%%%%%%%
%%%%%%%%%%%%%%%%%%%%%%%%%%%%%%%%%%%%%%%%%%%%


\section*{References}

\begin{thebibliography}{10}
%\begin{thebibliography}{}



%% Quantum memory general -- Intro %%


\bibitem{Phillips2001}
D. F. Phillips, A. Fleischhauer, A. Mair, R. L. Walsworth, 
and M. D. Lukin, 
Storage of Light in Atomic Vapor, 
Phys. Rev. Lett. {\bf 86}, 783 (2001). 

\bibitem{Hau2001}
C. Liu, Z. Dutton, C. H. Behroozi, and L. V. Hau, 
Observation of coherent optical information storage in an atomic medium using
halted light pulses, 
Nature \textbf{409}, 25, 2001. 

%\bibitem{Schori2002}
%C. Schori, B. Julsgaard, J. L. S{\o}rensen, and E. S. Polzik, 
%Recording quantum properties of light in a long-lived atomic 
%spin state : Towards quantum memory, 
%Phys. Rev. Lett. {\bf 89}, 057903 (2002). 

\bibitem{Polzik 2004}
B. Julsgaard, J. Sherson, J. I. Cirac, J. Fiurasek, and E. S. Polzik, 
Experimental demonstration of quantum memory for light, 
Nature \textbf{432}, 482, 2004. 

\bibitem{Kuzmich 2005}
T. Chaneliere, D. N. Matsukevich, S. D. Jenkins, S.-Y. Lan, T. A. B. Kennedy, 
and A. Kuzmich, 
Storage and retrieval of single photons transmitted between remote 
quantum memories, 
Nature \textbf{438}, 833/836, 2005. 

\bibitem{Sellars 2010}
M. Hedges, J. Longdell, Y. Li, and M. Sellars, 
Efficient quantum memory for light, 
Nature \textbf{465}, 1052/1056, 2010.



%% application of q memory to q repeater and communication %%

\bibitem{Repeater 1}
H. J. Briegel, W. Dur, J. I. Cirac, and P. Zoller, 
Quantum repeaters: The role of imperfect local operations in quantum 
communication, 
Phys. Rev. Lett. \textbf{81}, 26, 1998.

\bibitem{Duan 2001}
L. M. Duan, M. D. Lukin, J. I. Cirac, and P. Zoller, 
Long-distance quantum communication with atomic ensembles and 
linear optics, 
Nature \textbf{414}, 6862, 413/418, 2001.

\bibitem{Repeater review}
N. Sangouard, C. Simon, H. de Riedmatten, and N. Gisin, 
Quantum repeaters based on atomic ensembles and linear optics, 
Rev. Mod. Phys. \textbf{83}, 1, 2011.



%% Intro review %%

\bibitem{Lvovsky 2009}
A. I. Lvovsky, B. C. Sanders, and W. Tittel, 
Optical quantum memory, 
Nature Photonics \textbf{3}, 706, 2009.

\bibitem{Special issue}
Special issue on quantum memory, 
Ed. by J.-L. Le Gouet and S. A. Moiseev, 
J. Phys. B: At. Mol. Opt. Phys. \textbf{45}, 2012. 

\bibitem{Focus on}
Focus on quantum memory, 
Ed. by G. Brennen, E. Giacobino, and C. Simon, 
New J. Phys. 2013.



%% DF subsystem %%


\bibitem{KnightNJP2000}
A. Beige, D. Braun, and P. L. Knight, 
Driving atoms into decoherence-free states, 
New J. Phys. \textbf{2}, 22, 2000.

\bibitem{Wineland2001}
D. Kielpinski, V. Meyer, M. A. Rowe, C. A. Sackett, W. M. Itano, 
C. Monroe, and D. J. Wineland, 
A decoherence-free quantum memory using trapped ions, 
Science \textbf{291}, 1013, 2001. 

\bibitem{Lidar2003}
D. A. Lidar and K. B. Whaley, 
Decoherence-free subspaces and subsystems, 
Irreversible Quantum Dynamics, edited by F. Benatti and R. Floreanini, 
Lecture Notes in Physics \textbf{622}, 83, 
Springer Berlin/Heidelberg, 2003. 

\bibitem{BaconPRA2006}
D. Bacon, 
Operator quantum error-correcting subsystems for self-correcting 
quantum memories, 
Phys. Rev. A \textbf{73}, 012340, 2006. 



%% general memory schematic --- switching and long-living state %% 


\bibitem{Fleischhauer}
M. Fleischhauer and M. D. Lukin, 
Quantum memory for photons: Dark-state polaritons, 
Phys. Rev. A {\bf 65}, 022314, 2002. 

\bibitem{Chen PRL 2013}
Y.-H. Chen, et. al., 
Coherent optical memory with high storage efficiency and 
large fractional delay, 
Phys. Rev. Lett. \textbf{110}, 083601, 2013. 

\bibitem{LipsonNatPhys2007}
Q. Xu, P. Dong, and M. Lipson, 
Breaking the delay-bandwidth limit in a photonic structure, 
Nature Physics \textbf{3} 406, 2007.

\bibitem{Noda 2007}
Y. Tanaka, J. Upham, T. Nagashima, T. Sugiya, T. Asano, and S. Noda, 
Dynamic control of the Q factor in a photonic crystal nanocavity, 
Nature Mater. \textbf{6}, 862, 2007.

\bibitem{PrebleOptExpress2010}
A. W. Elshaari, A. Aboketaf, and S. F. Preble, 
Controlled storage of light in silicon cavities, 
Opt. Express \textbf{18}-3, 3014, 2010. 

\bibitem{Painter 2011}
D. E. Chang, A. H. Safavi-Naeini, M. Hafezi, and O. Painter, 
Slowing and stopping light using an optomechanical crystal array, 
New J. Phys. \textbf{13}, 023003, 2011.

\bibitem{Furusawa 2013}
J. Yoshikawa, K. Makino, S. Kurata, P. van Loock, and A. Furusawa, 
Creation, storage, and on-demand release of optical quantum states 
with a negative Wigner function, 
Phys. Rev. X \textbf{3}, 041028, 2013. 



%% No go theorem in DF memory %%

\bibitem{Kielpinski 2013}
D. Kielpinski, R. A. Briggs, and H. M. Wiseman, 
Unavoidable decoherence in the quantum control of an unknown state, 
Q. Meas. and Q. Metrology \textbf{1}, 1, 2013.






%% wave form optimization %% 

\bibitem{Gorshkov}
A. V. Gorshkov, A. Andre, M. D. Lukin, and A. S. Sorensen, 
Photon storage in Lambda-type optically dense atomic media, 
I. Cavity model, Phys. Rev. A \textbf{76}, 033804, 2007; 
II. Free-space model, Phys. Rev. A \textbf{76}, 033805, 2007; 
III. Effects of inhomogeneous broadening, Phys. Rev. A \textbf{76}, 033806, 2007.

\bibitem{Novikova 2007}
I. Novikova, A. V. Gorshkov, D. F. Phillips, A. S. Sorensen, M. D. Lukin, 
and R. L. Walsworth, 
Optimal control of light pulse storage and retrieval, 
Phys. Rev. Lett. \textbf{98}, 243602, 2007.

\bibitem{Novikova 2008}
I. Novikova, N. B. Phillips, and A. V. Gorshkov, 
Optimal light storage with full pulse-shape control, 
Phys. Rev. A \textbf{78}, 021802, 2008.

\bibitem{Novikova 2008 b}
N. B. Phillips, A. V. Gorshkov, and I. Novikova, 
Optimal light storage in atomic vapor, 
Phys. Rev. A \textbf{78}, 023801, 2008.



%% rising exponential %%

\bibitem{Muschik2006}
C. A. Muschik, K. Hammerer, E. S. Polzik, and J. I. Cirac, 
Efficient quantum memory and entanglement between light and an 
atomic ensemble using magnetic field, 
Phys. Rev. A \textbf{73}, 062329, 2006. 

\bibitem{He2009}
Q. Y. He, M. D. Reid, E. Giacobino, J. Cviklinski, and P. D. Drummond, 
Dynamical oscillator-cavity model for quantum memories, 
Phys. Rev. A \textbf{79}, 022310, 2009.

\bibitem{YiminWang2012}
Y. Wang, J. Minar, G. Hetet, and V. Scarani, 
Quantum memory with a single two-level atom in a half cavity, 
Phys. Rev. A \textbf{85}, 013823, 2012.

\bibitem{rising exp experiment 1}
S. A. Aljunid, G. Maslennikov, Y. Wang, H. L. Dao, V. Scarani, and C. Kurtsiefer, 
Excitation of a single atom with exponentially rising light pulses, 
Phys. Rev. Lett. \textbf{111}, 103001, 2013.

\bibitem{rising exp experiment 2}
M. Bader, S. Heugel, A. L. Chekhov, M. Sondermann, and G. Leuchs, 
Efficient coupling to an optical resonator by exploiting time-reversal symmetry, 
New J. Phys. \textbf{15}, 123008, 2013. 

\bibitem{rising exp experiment 3}
G. K. Gulati, B. Srivathsan, B. Chng, A. Cere, D. Matsukevich, and C. Kurtsiefer, 
Counterintuitive temporal shape of single photons, 
arXiv:1402.5800, 2014.




%%% linear quantum systems %%%


%%% linear optics

\bibitem{GardinerBook}
C. Gardiner and P. Zoller, 
{\it Quantum Noise}, 
Springer, Berlin, 2000. 

\bibitem{WisemanBook}
H. M. Wiseman and G. J. Milburn, 
{\it Quantum Measurement and Control}, 
Cambridge University Press, 2010.



%%% opto-mechanics

\bibitem{LawPRA1995}
C. K. Law, 
Interaction between a moving mirror and radiation pressure: 
A Hamiltonian formulation, 
Phys. Rev. A {\bf 51}, 2537, 1995. 

\bibitem{Chen 2013}
Y. Chen, 
Macroscopic quantum mechanics: theory and experimental concepts 
of optomechanics, 
J. Phys. B: At. Mol. Opt. Phys. \textbf{46}, 104001, 2013. 


%%% trapped particle

\bibitem{WinelandRMP2003}
D. Leibfried, R. Blatt, C. Monroe, and D. J. Wineland, 
Quantum dynamics of single trapped ions, 
Rev. Mod. Phys. \textbf{75}, 281, 2003. 

\bibitem{Polzik2011}
K. Jensen et. al., 
Quantum memory for entangled continuous-variable states, 
Nature Physics \textbf{7}, 13, 2011. 


%% atomic ensembles %%

\bibitem{Duan}
L. M. Duan, J. I. Cirac, and P. Zoller, 
Three-dimensional theory for interaction between atomic ensembles and 
free-space light, 
Phys. Rev. A \textbf{66}, 023818, 2002.

\bibitem{Kuzmich 2006}
D. N. Matsukevich, T. Chaneliere, S. D. Jenkins, S. Y. Lan, 
T. A. B. Kennedy, and A. Kuzmich, 
Deterministic single photons via conditional quantum evolution, 
Phys. Rev. Lett. \textbf{97}, 013601, 2006.

\bibitem{Parkins2006}
A. S. Parkins, E. Solano, and J. I. Cirac, 
Unconditional two-mode squeezing of separated atomic ensembles, 
Phys. Rev. Lett. \textbf{96}, 053602, 2006. 

\bibitem{Parkins2007}
F. Dimer, B. Estienne, A. S. Parkins, and H. J. Carmichael, 
Proposed realization of the Dicke-model quantum phase transition 
in an optical cavity QED system, 
Phys. Rev. A \textbf{75}, 013804, 2007.

\bibitem{Ficek2009}
G. Li, S. Ke, and Z. Ficek, 
Generation of pure continuous-variable entangled cluster states of four 
separate atomic ensembles in a ring cavity, 
Phys. Rev. A \textbf{79}, 033827, 2009. 

\bibitem{PolzikRMP2010}
K. Hammerer, A. S. Sorensen, and E. S. Polzik, 
Quantum interface between light and atomic ensembles, 
Rev. Mod. Phys. \textbf{82}, 1041, 2010. 



%% dark mode %%

\bibitem{Dong2012}
C. Dong, V. Fiore, M. C. Kuzyk, and H. Wang, 
Optomechanical dark mode, 
Science \textbf{338}, 1609, 2012.

\bibitem{Clerk2012}
Y. D. Wang and A. A. Clerk, 
Using dark modes for high-fidelity optomechanical quantum state transfer, 
New J. Phys. \textbf{14} 105010, 2012. 



%% control book -- zero dynamics %%

\bibitem{Zhou Doyle book} 
K. Zhou and J. C. Doyle, 
{\it Essentials of Robust Control}, 
Prentice Hall, 1997. 

\bibitem{Isidori}
A. Isidori, 
{\it Nonlinear Control Systems}, 3rd ed., 
Springer, 1995.

\bibitem{Nijmeijer}
H. Nijmeijer and A. van der Schaft, 
{\it Nonlinear Dynamical Control Systems}, 3rd ed., 
Springer, 1996.



%%% general passive %%%


\bibitem{GoughPRA2008}
J. E. Gough, R. Gohm, and M. Yanagisawa, 
Linear quantum feedback networks, 
Phys. Rev. A. \textbf{78}, 062104, 2008.

\bibitem{Guta Yamamoto}
M. Guta and N. Yamamoto, 
Systems identification for passive linear quantum systems:
the transfer function approach, 
arXiv:1303.3771; 
Proceedings of 52nd IEEE CDC, 2013.




%% single photon field %%

\bibitem{Zoller 1998}
K. M. Gheri, K. Ellinger, T. Pellizzari, and P. Zoller, 
Photon-wavepackets as flying quantum bits, 
Fortschr. Phys. \textbf{46}, 4-5, 401/415, 1998. 

\bibitem{Milburn2008}
G. J. Milburn, 
Coherent control of single photon states, 
Eur. Phys. J. \textbf{159}, 113/117, 2008.

\bibitem{Munro 2010}
W. J. Munro, K. Nemoto, G. J. Milburn, 
Intracavity weak nonlinear phase shifts with single photon driving, 
Optics Communications, \textbf{283}, 5, 741/746, 2010.

\bibitem{Baragiola}
B. Q. Baragiola, R. L. Cook, A. M. Branczyk, and J. Combes, 
N-photon wave packets interacting with an arbitrary quantum system
Phys. Rev. A \textbf{86}, 013811, 2012.

\bibitem{Guofeng2013}
G. Zhang and M. R. James, 
On the response of quantum linear systems to single photon 
input fields, 
IEEE Trans. Automat. Contr. \textbf{58}-5, 1221/1235, 2013. 



%%% case studies in infinite dim DF subsystems %%%

\bibitem{PrauznerJPA2004}
J. S. Prauzner-Bechcicki, 
Two-mode squeezed vacuum state coupled to the common thermal reservoir, 
J. Phys. A: Math. Gen. \textbf{37}, 173, 2004. 

\bibitem{Huang2013}
S. Huang, 
Double electromagnetically induced transparency and narrowing of 
probe absorption in a ring cavity with nanomechanical mirrors, 
%Optically dark state in a ring cavity with nanomechanical mirrors, 
J. Phys. B: At. Mol. Opt. Phys. \textbf{47}, 055504, 2014. 

\bibitem{Zambrini2013}
G. Manzano, F. Galve, and R. Zambrini, 
Avoiding dissipation in a system of three quantum harmonic oscillators, 
Phys. Rev. A \textbf{87}, 032114, 2013.

\bibitem{YamamotoDFS}
N. Yamamoto, 
Decoherence-free linear quantum subsystems, 
IEEE Trans. Automat. Contr. \textbf{59}-7, 1845/1857, 2014. 


%%% dark state principle %%%


\bibitem{Cirac 1997}
J. I. Cirac, P. Zoller, H. J. Kimble, and H. Mabuchi, 
Quantum state transfer and entanglement distribution among 
distant nodes in a quantum network, 
Phys. Rev. Lett. \textbf{78}, 3221, 1997. 

\bibitem{Beige 2000}
A. Beige, D. Braun, B. Tregenna, and P. L. Knight, 
Quantum computing using dissipation to remain in a decoherence-free 
subspace, 
Phys. Rev. Lett. \textbf{85}, 1762, 2000.


%%% Energy balance %%%

\bibitem{Hush}
M. R. Hush, A. R. R. Carvalho, M. Hedges, and M. R. James, 
Analysis of the operation of gradient echo memories using 
a quantum input-output model, 
New J. Phys. \textbf{15}, 085020, 2013.



%%% QEC %%%

\bibitem{Braunstein 1998}
S. L. Braunstein, 
Error correction for continuous quantum variables, 
Phys. Rev. Lett. \textbf{80}, 4084, 1998. 

\bibitem{Lloyd 1998}
S. Lloyd and J. J. E. Slotine, 
Analog quantum error correction, 
Phys. Rev. Lett. \textbf{80}, 4088, 1998. 


%%% non rising exp function %%%

\bibitem{Moiseev JPB 2012}
S. A. Moiseev, and S. N. Andrianov, 
Photon echo quantum random access memory integration 
in a quantum computer, 
J. Phys. B: At. Mol. Opt. Phys. \textbf{45}, 124017, 2012. 



%%% matching condition %%%


\bibitem{Afzelius 2010}
M. Afzelius and C. Simon, 
Impedance-matched cavity quantum memory, 
Phys. Rev. A \textbf{82}, 022310, 2010. 

\bibitem{Moiseev PRA 2010}
S. A. Moiseev, S. N. Andrianov, and F. F. Gubaidullin, 
Efficient multimode quantum memory based on photon echo 
in an optimal QED cavity, 
Phys. Rev. A \textbf{82}, 022311, 2010. 

\bibitem{Moiseev PRA 2013}
S. A. Moiseev, 
Off-resonant Raman-echo quantum memory for inhomogeneously 
broadened atoms in a cavity, 
Phys. Rev. A \textbf{88}, 012304, 2013.

\bibitem{Chaneliere 2014}
T. Chaneliere, 
Strong excitation of emitters in an impedance matched cavity: 
the area theorem, $\pi$-pulse and self-induced transparency, 
Optics Express \textbf{22}, 4423, 2014. 


\bibitem{Tittel 2010}
W. Tittel et. al., 
Photon-echo quantum memory in solid state systems, 
Laser and Photon. Rev. \textbf{4}-2, 244/267, 2010. 


%%% dark state principle Appendix %%%

\bibitem{Bouten2007}
L. Bouten, R. van Handel, and M. R. James, 
An introduction to quantum filtering, 
SIAM J. Contr. Optim. \textbf{46}-6, 2199/2241, 2007.

\bibitem{Gough 2012}
J. E. Gough, M. R. James, H. I. Nurdin, and J. Combes, 
Quantum filtering for systems driven by fields in single-photon states 
or superposition of coherent states, 
Phys. Rev. A \textbf{86}, 043819, 2012.


\end{thebibliography}
\end{document}